\documentclass[12pt]{article}
\pdfoutput=1
\usepackage[nosort]{cite}
\usepackage[height=8.85in,width=6.45in]{geometry}
\usepackage{pifont}

\usepackage{times}

\usepackage[utf8]{inputenc}
\usepackage{amsmath}
\usepackage{amssymb}
\usepackage{mathtools}
\numberwithin{equation}{section}
\usepackage{slashed}
\usepackage{braket}
\usepackage[svgnames,psnames]{xcolor}
\usepackage[colorlinks,citecolor=DarkGreen,linkcolor=FireBrick,linktocpage,pagebackref]{hyperref}
\usepackage{cite}
\usepackage{graphicx}
\usepackage{tikz}
\usepackage{tikz-cd}

\usepackage{courier}
\usepackage{bm}

\usepackage{dashbox}
\usepackage{subcaption}
\usepackage{enumitem}

\usepackage{upgreek}
\usepackage{mdframed}

\usepackage{blkarray}
\usepackage{arydshln}
\usepackage{dsfont}
\usetikzlibrary{patterns}
\usetikzlibrary{arrows.meta}

\usepackage{calc}

\usepackage{accents}

\newcommand{\ie}{\begin{equation}\begin{aligned}}
\newcommand{\fe}{\end{aligned}\end{equation}}

\renewcommand{\title}[1]{\vbox{\center\LARGE{#1}}\vspace{5mm}}
\renewcommand{\author}[1]{\vbox{\center#1}\vspace{5mm}}
\newcommand{\address}[1]{\vbox{\center\em#1}}

\newcommand*{\bS}{\mathbb{S}}
\newcommand*{\bT}{\mathbb{T}}

\begin{document}

\begin{titlepage}
 	\hfill YITP-SB-2022-28, MIT-CTP/5457
 	\\

\title{Non-invertible Time-reversal Symmetry}

\author{Yichul Choi${}^{1,2}$,   Ho Tat Lam${}^3$, and Shu-Heng Shao${}^1$}

		\address{${}^{1}$C.\ N.\ Yang Institute for Theoretical Physics, Stony Brook University\\
        ${}^{2}$Simons Center for Geometry and Physics, Stony Brook University\\
		${}^{3}$Center for Theoretical Physics, Massachusetts Institute of Technology
		}

\abstract
 
In gauge theory,  it is commonly stated that time-reversal symmetry only exists at $\theta=0$ or $\pi$ for a $2\pi$-periodic $\theta$-angle.  
In this paper, we point out that in both the free Maxwell theory and  massive QED, there is a non-invertible time-reversal symmetry at every rational $\theta$-angle, i.e., $\theta= \pi p/N$.   
The non-invertible time-reversal symmetry is  implemented by a conserved, anti-linear operator without an inverse. It  is a composition of the naive time-reversal transformation and a fractional quantum Hall state. 
We also find similar non-invertible time-reversal  symmetries  in non-Abelian gauge theories, including the $\mathcal{N}=4$ $SU(2)$ super Yang-Mills theory along the locus $|\tau|=1$ on the conformal manifold.

\end{titlepage}

\tableofcontents

\section{Introduction}

In gauge theory with a $2\pi$-periodic $\theta$-angle, there can be   a manifest time-reversal symmetry at $\theta=0$.  
At $\theta=\pi$, there is a slightly more subtle time-reversal symmetry, which is a composition of the naive time-reversal transformation with $\theta\mapsto \theta+2\pi$. 
In both cases, the time-reversal symmetry is  implemented by an anti-unitary operator $\mathsf{T}$  obeying $\mathsf{T}^\dagger \times \mathsf{T} =1$.  
The relation between $\mathsf{T}$ and $\mathsf{T}^\dagger$ depends on the details of the quantum system. Different time-reversal algebra, e.g., $\mathsf{T}^2=1, \mathsf{T}^2 =(-1)^F$, etc., can be used to place the quantum field theory (QFT) on manifolds with different tangential structures. 
In this paper, we discuss more subtle time-reversal symmetries at other values of $\theta$.

Our discussion follows closely the recent developments of a novel kind of generalized global symmetry, the \textit{non-invertible symmetry}. 
(See \cite{McGreevy:2022oyu,Cordova:2022ruw} for reviews of generalized global symmetries.) 
It is implemented by  conserved operators, or more generally,  topological defects \cite{Gaiotto:2014kfa} in a relativistic setting, that do not obey a group multiplication law. 
Yet, they are invariants under renormalization group flows and lead to nontrivial selection rules as well as constraints on the low-energy phase diagram. 
Recently, based on earlier works in 1+1 dimensions \cite{Verlinde:1988sn,Petkova:2000ip,Fuchs:2002cm,Frohlich:2004ef,Frohlich:2006ch,Frohlich:2009gb,Bhardwaj:2017xup,Chang:2018iay,Ji:2019ugf,Lin:2019hks,Thorngren:2019iar,Gaiotto:2020iye,Komargodski:2020mxz,Chang:2020imq,Thorngren:2021yso,Huang:2021zvu,Burbano:2021loy},  a large class of non-invertible symmetries was discovered in many familiar quantum systems in general spacetime dimensions \cite{Koide:2021zxj,Choi:2021kmx,Kaidi:2021xfk,Roumpedakis:2022aik,Bhardwaj:2022yxj,Cordova:2022rer,Arias-Tamargo:2022nlf,Hayashi:2022fkw,Choi:2022zal,Kaidi:2022uux,Choi:2022jqy,Cordova:2022ieu,Antinucci:2022eat,Bashmakov:2022jtl,Inamura:2022lun,Damia:2022rxw,Damia:2022bcd,Moradi:2022lqp,Chang:2022hud}, including the Standard Model \cite{Choi:2022jqy,Cordova:2022ieu}.\footnote{See also \cite{Tachikawa:2017gyf,Rudelius:2020orz,Heidenreich:2021xpr,Nguyen:2021yld,Kaidi:2021gbs,Wang:2021vki,Benini:2022hzx} for  discussions of non-invertible higher-form symmetries and  \cite{Ji:2019jhk,Kong:2020cie} for related discussions under the name of algebraic higher symmetry.}  
The key for constructing some of these symmetries is to compose a duality transformation (or more generally, an isomorphism of the quantum system) with the gauging of a discrete higher-form symmetry. 
See also \cite{Grimm:1992ni,Feiguin:2006ydp,Hauru:2015abi,Aasen:2016dop,Buican:2017rxc,Aasen:2020jwb,Inamura:2021szw,Koide:2021zxj,Huang:2021nvb,Vanhove:2021zop,Liu:2022qwn} for lattice realizations of non-invertible symmetries.

In this paper, we extend this construction to time-reversal symmetries. 
This was first discussed in \cite{Kaidi:2021xfk}, and we further generalize it to the free Maxwell theory, massive QED, ${\cal N}=4$ super Yang-Mills theory, and  invertible one-form symmetry-protected topological (SPT) phases.
 
In Maxwell theory  and QED, we find that for any rational $\theta$-angle,\footnote{More precisely, by a rational $\theta$-angle we mean $\theta/\pi$ is a rational number. We hope this slight abuse of terminology does not cause any confusion.} the composition of the naive time-reversal transformation and the gauging of  a magnetic one-form symmetry leaves the theory invariant. 
This leads to a conserved, anti-linear operator which is a composition of the naive time-reversal operator and a fractional quantum Hall state. 
However, because of the gauging operation that is involved, it is not an anti-unitary operator and it does not have an inverse -- it's a non-invertible time-reversal symmetry.  
More specifically, the product $\mathsf{T}^\dagger \times \mathsf{T} = {\cal C}\neq 1$ is a nontrivial \textit{condensation operator} \cite{Kong:2014qka,Else:2017yqj,Gaiotto:2019xmp,Kong:2020cie,Johnson-Freyd:2020twl,Roumpedakis:2022aik,Choi:2022zal} (see also \cite{Choi:2021kmx,Kaidi:2021xfk}), which plays a pivotal role in the recent developments of non-invertible symmetry.

It is surprising that abelian gauge theory is almost always time-reversal invariant in the space of the $\theta$-angle. 
Indeed,  the free Maxwell theory and QED has a non-invertible time-reversal symmetry for a dense subset   of the possible values for the $\theta$-angle, in addition to the invertible ones at $\theta=0,\pi$.  
This is perhaps related to the reason why the $\theta$-angle for an abelian gauge group is often overlooked, say, in the discussion of the strong $\mathsf{CP}$ problem. 
(Another reason for this is that there is no nontrivial instanton configuration in noncompact space for $U(1)$ because $\pi_3(U(1))=0$.)
Nonetheless, the $\theta$-angle of an abelian gauge theory (say the massive QED for the real world) is generally a physical  $\mathsf{CP}$-violating parameter in the Lagrangian that cannot be removed by any field redefinition.

\section{Free Maxwell theory}

Consider the free Maxwell theory of a dynamical $U(1)$ gauge field $A$ without any matter fields:
\begin{equation}
    \mathcal{L}_{\mathcal{Q}_\tau}= -\frac{1}{2e^2} F \wedge \star F + \frac{\theta}{8\pi^2} F \wedge F \,,
\end{equation}
where $F=dA$.\footnote{We will work in Lorentzian signature $(-+++)$ throughout the paper. We will use $t$ for the Lorentzian time coordinate. On the other hand, $\tau$ always stands for the complexified coupling  (rather than the Euclidean time coordinate).} 
The theory is parameterized by the complexified coupling constant
\begin{equation}
    \tau = \frac{\theta}{2\pi} + \frac{2\pi i}{e^2} \in \mathbb{H} \,,
\end{equation}
where $\mathbb{H}$ denotes the upper half-plane.
We will denote the free Maxwell theory at a given value of $\tau$ as $\mathcal{Q}_\tau$.\footnote{
    To completely characterize the Maxwell theory, one also needs to specify the spectrum of line operators and various quantum numbers associated with them.
    We will focus on the case where every line operator can be either bosonic or fermionic.
    Such a Maxwell theory can be defined consistently on spin manifolds with a choice of the spin structure, and it enjoys the full $SL(2,\mathbb{Z})$ duality.
     For the most part of the paper, we will assume that the spacetime manifold is  spin and in particular orientable.
    In this current paper, we will not attempt to define the Maxwell theory on general non-orientable manifolds using the invertible and non-invertible time-reversal symmetries that will be discussed in later sections.
}

The Maxwell theory has the $SL(2,\mathbb{Z})$ duality generated by $\bS$ and $\bT$. They act on the complexified coupling as
\begin{equation} \label{eq:S_T_operations}
    \bS: \,\tau \mapsto -\frac{1}{\tau} \,, \quad
    \bT: \, \tau \mapsto \tau +1 \,.
\end{equation}
$\bS$ acts on the local operators as
\ie\label{bSaction}
F\mapsto\frac{2\pi}{e^2}\star F-\frac{\theta}{2\pi}F~,
\fe 
and on the Wilson and 't Hooft line operators, denoted by $W_E$ and $W_M$, respectively, as
\ie
(W_E,W_M)\mapsto(W_M,W_E^\dagger)~.
\fe
Although $\bS^2 = -1 \in SL(2,\mathbb{Z})$ acts trivially on $\tau$, it acts non-trivially on operators as the $\mathbb{Z}_2^{(0)}$ charge conjugation symmetry $A\mapsto-A$, which exists at every value of $\tau$.\footnote{The superscript denotes the form degree of a global symmetry.} 
$\bT$ acts trivially on the local operators but non-trivially on the Wilson and 't Hooft line operators as
\ie
(W_E,W_M)\mapsto(W_E,W_M W_E^\dagger)~.
\fe
We have $(\bS \bT)^3 =1$.
Theories at different values of $\tau$ that are related by a duality transformation are identified. They form a fundamental domain of $SL(2,\mathbb{Z})$ in the upper half-plane,
\begin{equation}
    \mathcal{F} = \left\{
        \tau \in \mathbb{H} : |\tau|>1 , |\text{Re}(\tau)|<\frac{1}{2}
    \right\} \cup
    \mathcal{B}^{|\tau|=1} \cup \mathcal{B}^{\theta=\pi}
\end{equation}
where
\begin{align}
\begin{split}
    \mathcal{B}^{|\tau|=1} &= \left\{
        \tau \in \mathbb{H} : |\tau|=1 , 0\leq \text{Re}(\tau)\leq \frac{1}{2}
    \right\} \,, \\
    \mathcal{B}^{\theta=\theta_0} &=  \left\{
        \tau \in \mathbb{H} : |\tau|\geq 1 , \text{Re}(\tau)=\frac{\theta_0}{2\pi}
    \right\} \,.
\end{split}
\end{align}

At a generic value of the complexified coupling $\tau$, we define a naive time-reversal transformation
\begin{equation} \label{eq:K_operation}
    K:\, \tau \mapsto -\bar{\tau} \,, \quad t \mapsto -t \,,
\end{equation}
where $t$ is the (Lorentzian) time coordinate.  
We emphasize that $K$ is generally not a  symmetry of the Maxwell theory ${\cal Q}_\tau$ at a generic $\tau$. 
Rather, the $K$ transformation maps a Maxwell theory ${\cal Q}_\tau$ to its orientation reversal ${\cal Q}_{-\bar \tau}$; that is, it flips the sign of $\theta$.
We choose $K$ to act on the gauge field $A$ as follows
\begin{equation}
    K:\, A_0 (t,\vec{x}) \mapsto -A_0 (-t,\vec{x}) \,, \quad A_i (t,\vec{x}) \mapsto A_i (-t,\vec{x}) \,.
\end{equation}

\subsection{Invertible Time-Reversal Symmetries} \label{sec:inv_T_Maxwell}
 
We first review the invertible time-reversal symmetries of the Maxwell theory, which are well-known in the literature.
See, for instance, \cite{Seiberg:2016gmd}.
The Maxwell theory has an invertible time-reversal symmetry along  the following loci inside the fundamental domain $\mathcal{F}$:
\begin{equation}
    \tau \in \mathcal{B}^{\theta=0} \cup \mathcal{B}^{|\tau|=1} \cup \mathcal{B}^{\theta=\pi} \,: \, \text{invertible time-reversal.}
\end{equation}
The nature of the time-reversal symmetries at different loci are slightly different, as we discuss below: 
\begin{itemize}
\item
At $\theta=0$,  the points in $\mathcal{B}^{\theta=0} $ are left invariant under the naive time-reversal  transformation  $K$. Therefore, the naive  transformation $K$ is a  symmetry, which we denote as $\mathsf{T}^{\theta=0} $.
\item
At $\theta=\pi$, the points in $ \mathcal{B}^{\theta=\pi}$ are left invariant under the  $\bT K$ transformation.
So we have a time-reversal symmetry $\mathsf{T}^{\theta=\pi}$ associated with the $\bT K$ transformation.

\item
The points in $\mathcal{B}^{|\tau|=1}$ are invariant under the $\bS K$ transformation.
This is because the $\bS$ transformation  on ${\cal B}^{|\tau|=1}$ leaves the electric coupling constant $e$ invariant but flips the sign of the $\theta$-angle.
Therefore, we have a time-reversal symmetry $\mathsf{T}^{|\tau|=1}$ associated with the transformation $ \bS K$. 
\end{itemize}
All these time-reversal symmetries are invertible and square to the identity, $(\mathsf{T}^{\theta=0})^2 =(\mathsf{T}^{\theta=\pi})^2= (\mathsf{T}^{|\tau|=1})^2=1$. 
 
At special values of $\tau$, we have enhanced symmetries.
At $\tau = i \in \mathcal{B}^{\theta=0} \cap \mathcal{B}^{|\tau|=1}$, the theory has both $\mathsf{T}^{\theta=0}$ and $\mathsf{T}^{|\tau|=1}$ symmetries.
This is because the $\bS$ transformation becomes a $\mathbb{Z}_4^{(0)}$ internal symmetry at $\tau=i$, and the two time-reversal symmetries precisely differ by $\bS$.
The time-reversal symmetry $\mathsf{T}^{\theta=0}$ acts nontrivially on $\bS$ as $\mathsf{T}^{\theta=0} \bS \mathsf{T}^{\theta=0} = \bS^{-1}$.
This indicates that the symmetry group generated by the time-reversal and the $\bS$-duality at $\tau=i$ is the dihedral group of order 8,
\begin{equation}
    \tau=i :\, D_8^{\mathsf{T}} \,.
\end{equation}
The superscript $\mathsf{T}$ is to indicate that the symmetry group involves orientation-reversing elements.
On the other hand, at $\tau = e^{\pi i/3} \in \mathcal{B}^{|\tau|=1} \cap \mathcal{B}^{\theta=\pi}$, we have both $\mathsf{T}^{|\tau|=1}$ and $\mathsf{T}^{\theta=\pi}$ symmetries.
The two time-reversal symmetries differ by the $\bT \bS^{-1}$ transformation. Indeed, at $\tau = e^{\pi i/3}$, we have a $\mathbb{Z}_6^{(0)}$ internal symmetry generated by $\bT \bS^{-1}$ transformation.
Similar to before, the time-reversal symmetry $\mathsf{T}^{\theta=\pi}$ acts nontrivially on $\bT \bS^{-1}$ as $\mathsf{T}^{\theta=\pi}(\bT \bS^{-1})\mathsf{T}^{\theta=\pi} = (\bT \bS^{-1})^{-1}$.
Thus, they define the dihedral group of order 12,
\begin{equation} \label{eq:D12}
    \tau = e^{\pi i/3} : \, D_{12}^{\mathsf{T}} \,.
\end{equation}

We will make a final comment about the invertible time-reversal symmetry $\mathsf{T}^{|\tau|=1}$ along $|\tau|=1$.
The time-reversal symmetry at these points in the fundamental domain may look slightly unfamiliar compared to those at $\theta=0$ and  $\theta=\pi$. 
However, it is easy to see that under the $\bT \bS \bT$ transformation, the $|\tau|=1$ points get mapped to the points with $\theta=\pi$.
Thus,  the $\mathsf{T}^{|\tau|=1}$ time-reversal symmetry  is related to the familiar time-reversal symmetry at $\theta=\pi$ by a duality transformation.

\subsection{Non-invertible Time-Reversal Symmetries}\label{sec:maxwellnoninv}
 
We now turn to the new non-invertible time-reversal symmetries in Maxwell theory.  
We claim that for every rational value of the $\theta$-angle, 
\begin{align}
\theta= \pi p /N\,,
\end{align}
the Maxwell theory has a non-invertible time-reversal symmetry.
Here $N$ and $p$ are arbitrary integers satisfying $N>1$, $-N < p <N$, and gcd$(p,N)=1$.

We first note that there is a topological interface $\mathsf{I}_{2\pi p \over N}$ separating the Maxwell theory at $\theta$ and $\theta-2\pi p /N$ (both sharing the same electric coupling $e$). 
For simplicity, we start with $p=1$, then the  worldvolume Lagrangian was derived in \cite{Choi:2022jqy}:
\ie\label{CTSTLag}
\mathsf{I}_{2\pi \over N} \equiv \exp\left[  i \oint_M \left( {N\over 4\pi } a \wedge da +{1\over 2\pi }a\wedge dA \right)\right]\,.
\fe
Here $M$ is a three-manifold where the interface is supported on and $a$ is a dynamical one-form gauge field that only lives on $M$. 
This can be viewed as a $\nu=1/N$ fractional quantum Hall state on $M$ for the bulk electromagnetic gauge field $A$.
We omit the path integral over $a$ in the definition of the interface $\mathsf{I}_{2\pi \over N}$. 
To see that this is the correct topological interface, we insert it  along $M$: $x=0$, and assume that  the $\theta$-angles in the $x<0$ and $x>0$ regions are $\theta_-$ and $\theta_+$, respectively. 
The action for this defect configuration is
\ie
&\int_{x<0} \left( -{1\over 2e^2} F\wedge \star F   + {\theta_-\over 8\pi^2}F\wedge F\right)+\int_{x=0}  \left( {N\over 4\pi } a \wedge da +{1\over 2\pi }a\wedge dA \right)\\
&+\int_{x>0} \left( -{1\over 2e^2} F\wedge \star F   + {\theta_+\over 8\pi^2}F\wedge F\right)\,.
\fe
The equations of motion of $a$ on $M$ and the bulk electromagnetic gauge field $A$ respectively give
\ie
&Nda + F=0 \,,\\
&{1\over 2\pi}(\theta_+-\theta_-) F= da\,.
\fe
Combining the two equations, we find $\theta_+ - \theta_- =  - 2\pi /N$, so \eqref{CTSTLag} is indeed the worldvolume Lagrangian for the topological interface separating the Maxwell theory with $\theta$ and $\theta - 2\pi /N$. 

Starting at $\theta= \pi  /N$, we can compose the naive time-reversal interface $K$ (which separates a QFT with its orientation-reversal) with the topological interface $\mathsf{I}_{2\pi \over N}$, which maps the $\theta$-angle back to itself:
\begin{equation}\label{composition}
    \theta =\pi /N~ \xmapsto{~~\mathsf{I}_{2\pi \over N}~~}~ \theta = -\pi /N~ \xmapsto{~~K~~} ~\theta = \pi /N \,. 
\end{equation}
We therefore conclude that 
\ie
\mathsf{T}^{\theta={\pi \over N}} \equiv K \circ \mathsf{I}_{2\pi \over N}
\fe
is an orientation-reversing, topological defect in  Maxwell theory at $\theta=\pi /N$. 
When we choose $M$ to be the whole space at a given time,  $\mathsf{T}^{\theta={\pi \over N}}$ becomes an anti-linear, conserved  operator that flips the sign of the Lorentzian time. 
Thus, it is a time-reversal symmetry. 
Intuitively, $\mathsf{T}^{2\pi \over N}$ is a composition of the naive time-reversal transformation and a fractional quantum Hall state \eqref{CTSTLag}.

Interestingly, the anti-linear operator $\mathsf{T}^{\theta={ \pi \over N}}$ is not anti-unitary. 
To see this, we compute 
\begin{align} \label{eq:fusion}
\begin{split}
& (\mathsf{T}^{\theta={ \pi \over N}} )^\dagger\times \mathsf{T}^{\theta={ \pi \over N}} 
=( \mathsf{I}_{2\pi \over N}) ^\dagger \times \mathsf{I}_{2\pi \over N}\\
&
= \exp \left[        i\oint_M \left(    \frac{N}{4\pi} ada  - {N\over 4\pi} \bar ad\bar a +\frac{1}{2\pi}( a-\bar a)dA
        \right)
    \right] \equiv {\cal C}^{(N)}\,.
    \end{split}
\end{align}
Here $(\mathsf{I}_{2\pi \over N})^\dagger  $ is the orientation-reversal of the interface $\mathsf{I}_{2\pi \over N}$:
\ie
(\mathsf{I}_{2\pi \over N})^\dagger    =  \mathsf{I}_{- {2\pi \over N}}=  \exp\left[  -i \oint_M \left( {N\over 4\pi }\bar a \wedge d\bar a +{1\over 2\pi }\bar a\wedge dA \right)\right]
\fe
where $\bar a$ is the dynamical one-form gauge field living on $(\mathsf{I}_{2\pi \over N})^\dagger $. 
The operator ${\cal C}^{(N)}$ is known as the \textit{condensation operator/defect} (see \cite{Roumpedakis:2022aik,Choi:2022zal,Choi:2022jqy} and Appendix \ref{app:condensation} for more details). 
The important point is that $\mathcal{C}^{(N)}$ is not a trivial operator; it acts nontrivially on the 't Hooft lines \cite{Choi:2022jqy}.
Thus, the time-reversal symmetry ${\mathsf{T}}^{\theta=\frac{\pi }{N}}$ is non-invertible, and, in particular, is not implemented by an anti-unitary operator.

For $\theta=\pi p /N$ at a more general $p$, we generalize the topological interface \eqref{CTSTLag} to
\ie\label{CTSTLagp}
\mathsf{I}_{2\pi p\over N} \equiv \exp\left[   i\oint_M  {\cal A}^{N,p} [dA/N]\right]\,,
\fe
where ${\cal A}^{N,p}[B]$ is the 2+1d minimal $\mathbb{Z}_N$ TQFT  \cite{Hsin:2018vcg} (see also \cite{Moore:1988qv,Bonderson:2007ci,Barkeshli:2014cna}) coupled to the $\mathbb{Z}_N^{(1)}$ background two-form gauge field $B$. 
It describes a $\nu = p/N$ fractional quantum Hall state. 
The non-invertible time-reversal symmetry at $\theta=\pi p/N$ is defined as 
\ie \label{eq:noninv_time_reversal}
\mathsf{T}^{\theta= {\pi p\over N}} \equiv K\circ \mathsf{I}_{2\pi p\over N}\,.
\fe 
Following a similar calculation in \cite{Choi:2022jqy}, we find that it obeys the non-invertible fusion rule  
\ie\label{fusion}
( \mathsf{T}^{\theta= {\pi p\over N}})^\dagger \times \mathsf{T}^{\theta= {\pi p\over N}} = {\cal C}^{(N)}\,.
\fe

At the intersection between $\theta=\pi p/N$ and $|\tau|=1$, i.e., at $\tau  = {p\over 2N} + i \sqrt{ 1- {p^2\over 4N^2}}$, there is an additional non-invertible, linear symmetry.
This comes from the composition of the $\mathsf{I}_{\frac{2\pi p}{N}}$ interface with the $\mathbb{S}$-duality interface (instead of the $K$ interface). 
The $\mathbb{S}$-duality interface has the same effect of flipping the sign of the $\theta$-angle when $|\tau|=1$ as we discussed earlier.
The worldvolume Lagrangian for this non-invertible, linear symmetry defect is obtained by composing the worldvolume Lagrangian \eqref{CTSTLagp} for the $\mathsf{I}_{\frac{2\pi p}{N}}$ interface with that for the $\mathbb{S}$ transformation which was discussed in \cite{Gaiotto:2008ak,Kapustin:2009av}:
\begin{equation}\label{Nality}
    \mathbb{S} \circ \mathsf{I}_{\frac{2\pi p}{N}}
    = \exp\left[ i\oint_M    \left(
    {\cal A}^{N,p} [dA_L/N] + \frac{1}{2\pi} A_L \wedge dA_R
    \right)    \right] \,,
\end{equation}
where $A_L$ and $A_R$ denote the bulk gauge fields to the left and to the right of the defect, respectively.
In addition, we have the orientation-reversal  of the defect \eqref{Nality}, which is obtained by taking the complex conjugation in \eqref{Nality} and exchanging $A_L\leftrightarrow A_R$. 
When $p=1$, it is straightforward to verify that the equations of motion are consistent.\footnote{
 By solving the equations of motion in the presence of the defect \eqref{Nality}, we can deduce that the defect acts on the field strength as $F_L \mapsto F_R= \frac{p}{2N}F_L + \sqrt{1-\frac{p^2}{4N^2}} \star F_L$,  which also follows from the $\mathbb{S}$-duality action \eqref{bSaction} on the field strength at the shifted value of $\tau = - {p\over 2N} +i \sqrt{1- {p^2\over 4N^2}}$.} 

We conclude that the free Maxwell theory is time-reversal invariant at $\theta= \pi p/N$. 
The notion of naturalness \cite{tHooft:1979rat} becomes rather exotic as we vary the $\theta$-angle: at every rational $\theta$-angle, there is a different non-invertible time-reversal symmetry. The situation is somewhat similar to the 1+1d compact boson CFT, where at every rational radius square $R^2$ there is a different enhanced chiral algebra.

\subsection{Gauging Magnetic One-form Symmetries}\label{sec:gauging}

Here we give an alternative construction for the non-invertible time-reversal symmetries by  gauging a discrete magnetic symmetry. 
This construction gives a rigorous proof why the non-invertible time-reversal symmetry is topological, and in particular, conserved.

The Maxwell theory has a magnetic $U(1)^{(1)}$ one-form symmetry, whose conserved two-form current is $J^{(m)} = F/2\pi$ obeying the conservation equation $dJ^{(m)}=0$ \cite{Gaiotto:2014kfa}. 
The Lagrangian of the Maxwell theory $\mathcal{Q}_\tau$ in the presence of the background two-form gauge field $B$ for the $U(1)^{(1)}$ magnetic one-form symmetry is given by
\begin{equation}
    \mathcal{L}_{\mathcal{Q}_\tau}[B] = -\frac{1}{2e^2} F \wedge \star F + \frac{\theta}{8\pi^2} F \wedge F 
    + \frac{1}{2\pi} F \wedge B\,.
\end{equation}
From now on until the end of this subsection, we assume $\theta= \pi p/N$.

Let $k$ be the multiplicative inverse of $p$ modulo $N$, i.e., $pk=1$ mod $N$. 
At $\theta= \pi p/N$,  we define the following transformations associated with the $\mathbb{Z}_N^{(1)}$ subgroup of the magnetic $U(1)^{(1)}$ one-form symmetry:\footnote{In \cite{Choi:2021kmx,Choi:2022zal}, $S,T$ are defined by gauging the electric one-form symmetries of the Maxwell theory, rather than the magnetic ones. 
}
\begin{align} \label{eq:sigma_tau_operations}
    \begin{split}
        S &: \quad \mathcal{L}_{S\mathcal{Q}_\tau}[B] = \mathcal{L}_{\mathcal{Q}_\tau}[b]
        +\frac{N}{2\pi} b \wedge dc 
        + \frac{Nk}{2\pi} b \wedge B \,, \\
        T &: \quad \mathcal{L}_{T \mathcal{Q}_\tau}[B] = \mathcal{L}_{\mathcal{Q}_\tau}[B]
        +\frac{Nk}{4\pi} B \wedge B \,, \\
        C &: \quad \mathcal{L}_{C\mathcal{Q}_\tau}[B] = \mathcal{L}_{\mathcal{Q}_\tau}[-B] \,.
    \end{split}
\end{align}
(To avoid cluttering, we have suppressed the $k,p,N$ dependence of the transformations $S,T,C$.) 
Note that $k$ and $k+N$ are identified, $k \sim k+N$.
Here $B$ is restricted to be a $\mathbb{Z}_N$-valued two-form gauge field.
The $S$ transformation corresponds to gauging the $\mathbb{Z}_N^{(1)} \subset U(1)^{(1)}$ subgroup of the magnetic one-form symmetry, where $b$ and $c$ are dynamical $U(1)$ two-form and one-form gauge fields, respectively. 
The integer $k$ specifies the generator of the dual $\mathbb{Z}_N^{(1)}$ one-form symmetry \cite{Gaiotto:2014kfa} that we obtain after gauging.
The $T$ transformation corresponds to stacking a $\mathbb{Z}_N^{(1)}$-SPT phase.  
Finally, $C=S^2$ flips the sign of the coupling to the background gauge field $B$. 
These transformations define a (projective) $SL(2,\mathbb{Z}_N)$ action  on the Maxwell theories \cite{Gaiotto:2014kfa,Bhardwaj:2020ymp,Choi:2022jqy}.

In \cite{Choi:2021kmx,Choi:2022zal} (see also \cite{Hayashi:2022fkw}), the $SL(2,\mathbb{Z})$ duality transformations $\bS,\bT$ are combined with the electric gauging operations $S,T$  to construct linear, non-invertible symmetries in the free Maxwell theory. At $\tau=iN$, there is a non-invertible duality defect ${\cal D} $ associated with the transformation ${\bS}S$. 
At $\tau = N e^{2\pi i /3}$, there is a triality defect associated with $\mathbb{S} ST$ for even $N$, and with $\mathbb{S}\mathbb{T}^{-1} ST$ for odd $N$. 
On these special points, which all have $\theta=0,\pi$ when brought into the fundamental domain, one can compose these duality and triality defects with the invertible time-reversal symmetries to obtain anti-linear, non-invertible symmetries.

In addition to the above symmetries, we will compose the magnetic $S,T,C$ operations with the naive time-reversal transformation $K$ to construct more general anti-linear, non-invertible symmetries away from $\theta=0,\pi$. 
 In the presence of the background gauge field $B$, we choose $K$ to act on the latter as
\ie
&  B_{0i}(t,\vec{x})\mapsto +B_{0i}(-t,\vec{x}),~~~ B_{ij}(t,\vec{x})\mapsto -B_{ij}(-t,\vec{x})\,.
\fe
Using \eqref{eq:sigma_tau_operations}, we find that the transformations $S,T,C$ obey the following commutation relations with the naive time-reversal transform $K$:  $K T = T^{-1} K \,,     K S = S^{-1} K \,,$ and   $ K C = C K$.

The essential idea to construct the non-invertible time-reversal symmetry is that we can compensate the action of $K$, which flips the sign of the $\theta$-angle, by using a certain discrete gauging if the value of the $\theta$-angle is rational.
In particular, it was shown in \cite{Choi:2022jqy} that the following transformation on the Maxwell theory has the effect of shifting the $\theta$-angle by a rational number:
\begin{equation}
CTST \mathcal{Q}_\tau = \mathcal{Q}_{\tau - \frac{p}{N}} \,.
\end{equation}
Thus, if we start from $\theta = \pi p/N$,  the composite transformation $KCT  S T=  CT^{-1}KST$ leaves the Maxwell theory at $\theta = \pi p/N$ invariant.\footnote{
    Equivalently, if we redefine the Maxwell theory to be $\mathcal{Q}'_\tau = T \mathcal{Q}_\tau$, then $\mathcal{Q}'_\tau$ has the non-invertible $CK S$ time-reversal symmetry at rational $\theta$-angles.
}
By performing this sequence of transformations $CT^{-1} K S T$ in half of the spacetime, this defines an anti-linear symmetry in Maxwell theory at $\theta=\pi p /N$, which we denote by $\mathsf{T}^{\theta=\frac{\pi p}{N}}$. 
To see that  this definition via gauging coincides with the one given in \eqref{eq:noninv_time_reversal}, we note that it was shown in \cite{Choi:2022jqy} that the interface introduced by the $CTST$ gauging is precisely \eqref{CTSTLagp}.
A similar non-invertible time-reversal symmetry in non-abelian gauge theory was first discussed in \cite{Kaidi:2021xfk}, which we will return to in Section \ref{sec:N=4} and  Appendix \ref{app:PSUN}. 
We summarize the time-reversal symmetries in the Maxwell theory   in Table \ref{table:T}.

The non-invertible fusion rule \eqref{fusion} can also be derived from the gauging perspective. 
By composing $CT^{-1} K S T$ with its inverse, we are left with the fusion between the interface associated with $ST$ and its orientation-reversal.
Importantly, while the $T$ transformation leads to an invertible topological interface between two QFTs, the interface associated with the $ST$ transformation (and similarly the one associated with $S$) is not invertible \cite{Choi:2021kmx,Kaidi:2021xfk}.  
In fact, the composition of the interface associated with $ST$ with its orientation-reversal is precisely the condensation defect ${\cal C}^{(N)}$ \cite{Choi:2022jqy}, which arises from the one-gauging  \cite{Roumpedakis:2022aik} of the $\mathbb{Z}_N^{(1)}$ magnetic one-form symmetry with a particular twist along a codimension-one manifold.

 \begin{table} 
    \centering
    \begin{tabular}{|c|c|c|} 
        \hline
~complexified coupling ~$\tau$~& ~time-reversal symmetry~ $\mathsf{T}$~  &~~ invertible?  ~~\\
      \hline
        $\theta=0$ & $K$   & Yes\\
        \hline
                $\theta=\pi $ & $\mathbb{T} K$   & Yes\\
                \hline
                 $|\tau|=1$ & $\mathbb{S} K$   & Yes\\
\hline
$\theta= \pi p/N$  &  ~~$CT^{-1} KST$~~  & No\\
        \hline
    \end{tabular}
    \caption{Summary of time-reversal symmetries and their associated transformations on the conformal manifold of the free Maxwell gauge theory. Here $K$ is the naive time-reversal transformation, $\mathbb{S},\mathbb{T}$ generate the $SL(2,\mathbb{Z})$ electromagnetic duality, and $S,T,C$ generate a (projective) $SL(2,\mathbb{Z}_N)$ action via gauging the magnetic $\mathbb{Z}_N^{(1)}$ one-form symmetry.  For the invertible time-reversal symmetries, we have omitted transformations that  only act on the couplings to the background gauge fields. 
    }
    \label{table:T}
\end{table}

There is an enhanced symmetry at the intersection point between ${\cal B}^{\theta = {\pi p\over N}}$ and ${\cal B}^{|\tau|=1}$. 
At this point, we have both the invertible and non-invertible time reversal symmetries, $\mathsf{T}^{|\tau|=1} $ (which is associated with  $\mathbb{S}K$) and $\mathsf{T}^{\theta=  {\pi p\over N}}$ (which is associated with $CT^{-1} KST$), respectively. 
Composing the two anti-linear symmetries, we obtain a non-invertible, linear symmetry at $\tau  = {p\over 2N} + i \sqrt{ 1- {p^2\over 4N^2}}$ associated with the $\mathbb{S}C TST$ transformation, which is implemented by the  defect \eqref{Nality}. 
Indeed, as was explained there, this defect is the composition of the $\mathbb{S}$-duality interface with the $\mathsf{I}_{\frac{2\pi p}{N}}$ interface, where the $\mathsf{I}_{\frac{2\pi p}{N}}$ interface implements the $CTST$ transformation.

\begin{figure}[!t]
    \centering
    \includegraphics[width=0.7\textwidth]{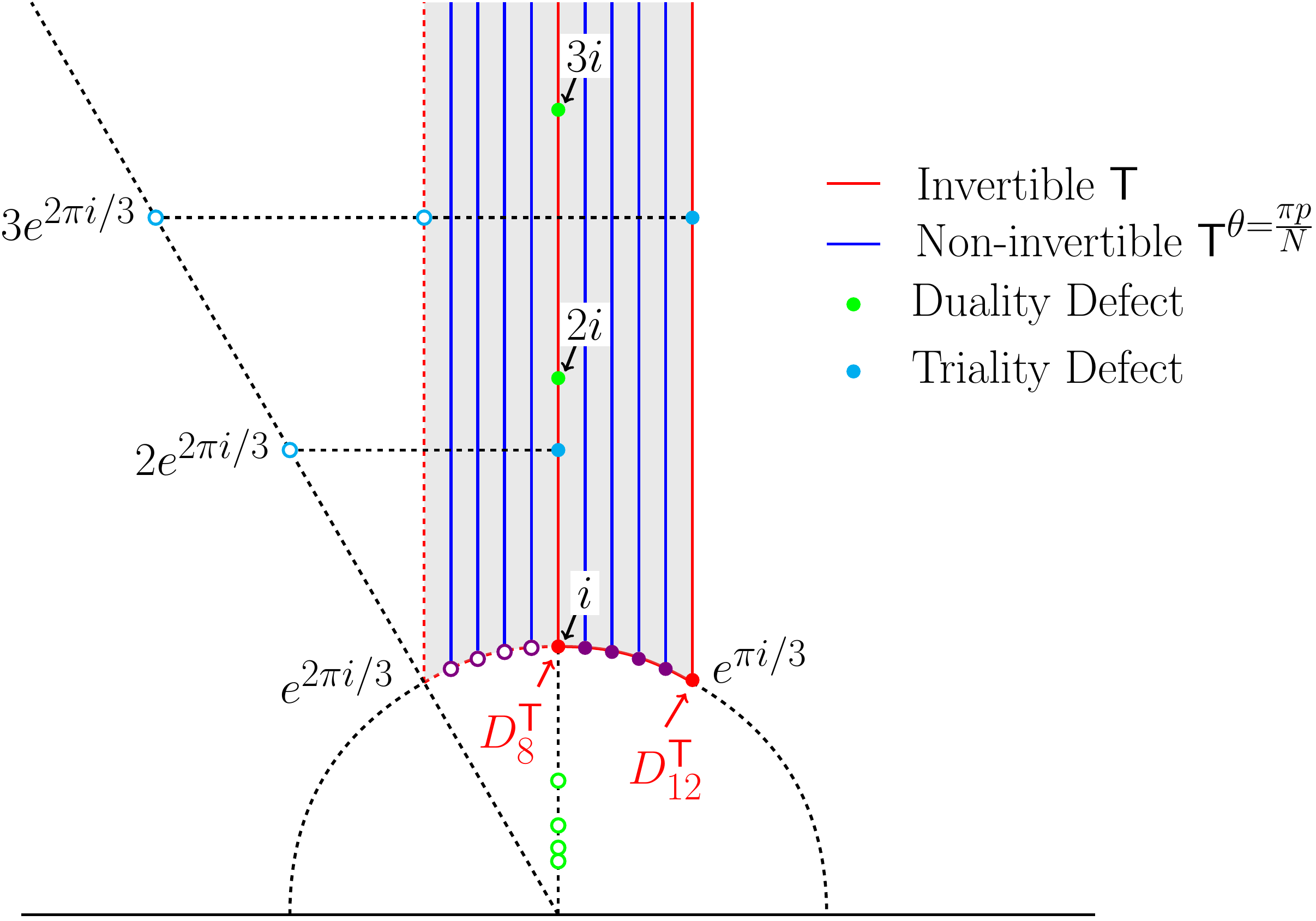}
    \caption{Some (anti-)linear (non-)invertible   symmetries of the Maxwell theory.
    The locus where the theory has an invertible time-reversal symmetry is indicated by red lines.
    At $\tau =i$ and at $\tau = e^{i\pi/3}$, we have enhanced symmetries $D_8^{\mathsf{T}}$ and $D_{12}^{\mathsf{T}}$, respectively.
    At every rational value of the $\theta$-angle with $\theta=\pi p/N$, there is a non-invertible time-reversal symmetry $\mathsf{T}^{\theta=\frac{\pi p}{N}}$, which is indicated by the vertical blue lines.
    When a blue line intersects with the $|\tau|=1$ locus at a purple dot, the non-invertible time-reversal symmetry factorizes into the invertible time-reversal symmetry $\mathsf{T}^{|\tau|=1}$ and a linear, non-invertible  symmetry \eqref{Nality}. 
    Some of the linear non-invertible symmetries of the Maxwell theory  are also indicated.
    At the   green dots, i.e., at $\tau = iN$, the theory realizes a linear, non-invertible duality defect  \cite{Choi:2021kmx}. 
    At the  cyan dots, i.e., at $\tau = Ne^{2\pi i/3}$, the theory realizes a linear, non-invertible triality defect \cite{Choi:2022zal}.  
    At these green/cyan points, there are non-invertible time-reversal symmetries obtained by composing the duality/triality defects with the invertible time-reversal symmetries. 
    Finally, there is a $\mathbb{Z}_2^{(0)}$ charge conjugation symmetry and a $U(1)^{(1)}\times U(1)^{(1)}$ one-form symmetry everywhere. 
    The red dashed lines and the hollow dots are outside the fundamental domain $\cal F$ and are related to the solid redlines and the solid dots, respectively, by duality transformations.}
    \label{Fig:fundamental_domain}
\end{figure}

In Figure~\ref{Fig:fundamental_domain}, we illustrate the (non-)invertible time-reversal symmetries and some of the linear non-invertible symmetries of the Maxwell theory across the fundamental domain $\mathcal{F}$, as well as various symmetry enhancements at special values of $\tau$.

\subsection{Mixed Anomalies at $\theta\in \pi \mathbb{Z}$}

We now discuss various mixed anomalies between the invertible time-reversal symmetry at $\theta\in\pi \mathbb{Z}$ and the one-form symmetries in the Maxwell theory.
For even $N$, we will find that the non-invertible time-reversal symmetry  at $\theta= \pi p /N$ arises from this mixed anomaly via gauging, generalizing the construction in \cite{Kaidi:2021xfk}. In contrast, for odd $N$, there is no corresponding anomaly and the non-invertible time-reversal symmetry has a more subtle origin.

\subsubsection*{Anomaly in $U(1)^{(1)}\times U(1)^{(1)}$}

First, it is well-known that there is mixed anomaly between the $U(1)^{(1)}$ electric and $U(1)^{(1)}$ magnetic one-form symmetries \cite{Gaiotto:2014kfa}. 
The Lagrangian of the Maxwell theory $\mathcal{Q}_\tau$ in the presence of the electric and magnetic two-form background gauge fields $C$ and $B$ is given by
\ie
\mathcal{L}_{\mathcal{Q}_\tau}[B,C]=-\frac{1}{2e^2}(F-C)\wedge \star(F-C)+\frac{\theta}{8\pi^2}(F-C)\wedge (F-C)+\frac{1}{2\pi}(F-C)\wedge B~.
\fe
The background one-form gauge symmetries act as
\ie
B\sim B+d\lambda~,\quad C\sim C+d\rho~,\quad A\sim A+\rho~.
\fe
They transform the Lagrangian anomalously as
\ie
\mathcal{L}_{\mathcal{Q}_\tau}[B,C]\mapsto \mathcal{L}_{\mathcal{Q}_\tau}[B,C]-\frac{1}{2\pi}C\wedge d\lambda~,
\fe
where we omit total derivatives that integrate to integer multiples of $2\pi$.
To preserve the background gauge symmetry, we can extend the background gauge fields $B$ and $C$ to a five-dimensional bulk with an invertible field theory described by the classical Lagrangian
\ie\label{eq:inv_field_theory}
\mathcal{L}_{\text{Inv}}[B,C]=\frac{1}{2\pi}B\wedge dC~.
\fe
Since the invertible field theory is nontrivial on closed five-dimensional manifolds, this represents a mixed anomaly between the $U(1)^{(1)}$ electric and $U(1)^{(1)}$ magnetic one-form symmetry, meaning that the anomalous transformation of the Lagrangian cannot be fixed by adding a four-dimensional counterterm.

\subsubsection*{Mixed Anomaly between $\mathsf{T}$ and $U(1)^{(1)}\times U(1)^{(1)}$}

Next, we move on to discuss a mixed anomaly between the time-reversal symmetry $\mathsf{T}^{\theta=\pi q}=\bT^q K$ and the $U(1)^{(1)}\times U(1)^{(1)}$ one-form symmetry at $\theta=\pi q$. 
Since we will turn on nontrivial background gauge fields for the $U(1)^{(1)}\times U(1)^{(1)}$ one-form symmetries, the $\theta$-angle is no longer $2\pi$-periodic, and we will work with $\theta=\pi q$  with a general integer $q$.

At $\theta=\pi q$, the time-reversal symmetry $\mathsf{T}^{\theta=\pi q}$ acts on the background gauge fields as
\ie\label{eq:TR_on_B_C}
&C_{0i}(t,\vec{x})\mapsto -C_{0i}(-t,\vec{x})~,\quad  B_{0i}(t,\vec{x})\mapsto +B_{0i}(-t,\vec{x})-qC_{0i}(-t,\vec{x})~,
\\
&C_{ij}(t,\vec{x})\mapsto +C_{ij}(-t,\vec{x})~,\quad B_{ij}(t,\vec{x})\mapsto -B_{ij}(-t,\vec{x})+qC_{ij}(-t,\vec{x})~.
\fe
In the presence of the background gauge fields $B$ and $C$, the Lagrangian transforms anomalously under the time-reversal action:
\ie\label{eq:anomaly_TR}
\mathcal{L}_{\mathcal{Q}_\tau}[B,C]\mapsto\mathcal{L}_{\mathcal{Q}_\tau}[B,C]+\frac{q}{4\pi}C\wedge C~.
\fe
This signals a mixed anomaly between $\mathsf{T}^{\theta= \pi q}$ and $U(1)^{(1)}\times U(1)^{(1)}$.

The anomalous time-reversal transformation \eqref{eq:anomaly_TR} can be canceled by  a five-dimensional invertible field theory. 
It turns out that on five-dimensional spin oriented manifolds, it is the same invertible field theory \eqref{eq:inv_field_theory} that cancels the mixed anomaly between $U(1)^{(1)}$ electric and $U(1)^{(1)}$ magnetic one-form symmetry.\footnote{On unoriented manifolds, there are additional terms involving the first Stiefel-Whitney class $w_1$ of the manifolds.} The time-reversal transformation \eqref{eq:TR_on_B_C}  shifts the Lagrangian of the invertible field theory by a total derivative
\ie
\mathcal{L}_{\text{Inv}}[B,C]\mapsto \mathcal{L}_{\text{Inv}}[B,C]-\frac{q}{4\pi}d(C\wedge C)~,
\fe
which exactly cancels the anomalous time-reversal transformation of $\mathcal{L}_{Q_\tau}[B,C]$ on the boundary.

\subsubsection*{Mixed Anomaly between $\mathsf{T}$ and $\mathbb{Z}_N^{(1)}$ }

We can restrict the $U(1)^{(1)}\times U(1)^{(1)}$ background gauge fields, $B$ and $C$, to the background gauge field for the $\mathbb{Z}_N^{(1)}$ electric one-form symmetry. Since the time-reversal symmetry acts non-trivially on the background gauge fields, to preserve the latter, we can turn off the magnetic background gauge field $B$ only when $q=pN$ with an integer $p$. At $\theta=\pi pN$, the anomalous transformation \eqref{eq:anomaly_TR} then reduces to
\ie\label{eq:anomaly_TRZN}
\mathcal{L}_{\mathcal{Q}_\tau}[0,C]\mapsto\mathcal{L}_{\mathcal{Q}_\tau}[0,C]+\frac{pN}{4\pi}C\wedge C
~.
\fe
where $C$ is restricted to be a $\mathbb{Z}_N$-valued two-form gauge field.

For even $N$,  there is no 3+1d counterterm that cancels the anomalous transformation \eqref{eq:anomaly_TRZN}. 
In contrast, when $N$ is odd, the anomalous transformation can be fixed by adding a classical counterterm to the Lagrangian
\ie\label{eq:counterterm}
\hat{\mathcal{L}}_{\mathcal{Q}_\tau}[0,C]=\mathcal{L}_{\mathcal{Q}_\tau}[0,C]+\frac{pN}{4\pi}\frac{N+1}{2} C\wedge C
~.
\fe
The new Lagrangian $\hat{\mathcal{L}}_{\mathcal{Q}_\tau}[C]$ is invariant under the time-reversal symmetry modulo terms that integrate to integer multiples of $2\pi$.

At $\theta= \pi pN$, we conclude that for even $N$, there is a mixed anomaly between the invertible time-reversal symmetry $\mathsf{T}^{\theta=\pi pN}$ and the electric $\mathbb{Z}_N^{(1)}$ one-form symmetry, while for odd $N$ there is no anomaly.\footnote{This anomaly can be embedded into non-Abelian gauge theory as follows. Consider for example the case $N=2$ and $p=1$. The $SU(2)$ pure Yang-Mills theory at $\theta_{UV}=\pi$ has a mixed anomaly between $\mathsf{T}$ and the $\mathbb{Z}_2^{(1)}$ center one-form symmetry \cite{Gaiotto:2017yup}. We can couple the $SU(2)$ gauge field to an adjoint scalar and then turn on a vev to Higgs the gauge group to $U(1)$. The UV $\theta$-gangle is related to the IR one as $\theta_{IR} =2 \theta_{UV} $. Hence the  anomaly in the $SU(2)$ theory is matched by that in the free $U(1)$ Maxwell theory at $\theta_{IR}=2\pi$. 
}

\subsubsection*{Non-invertible Time-reversal Symmetry from Gauging}

What happens if we gauge the $\mathbb{Z}_N^{(1)}$ electric one-form symmetry at $\theta=\pi pN$? Recall that there are multiple ways to gauge a $\mathbb{Z}_N^{(1)}$ one-form symmetry. We will restrict to the simplest case where we simply promote the background gauge field to a dynamical gauge field without stacking a $\mathbb{Z}_N^{(1)}$-SPT phase. The Lagrangian after gauging is
\ie
\mathcal{L}_{{\cal Q}_{\tau/N^2}}[B,0]=\mathcal{L}_{\mathcal{Q}_\tau}[0,c]+\frac{N}{2\pi} c\wedge du+\frac{N}{2\pi}c\wedge B 
~,
\fe
where  $c$ is a dynamical $U(1)$ two-form gauge field and $u$ is a dynamical $U(1)$ one-form gauge field that constrains $c$ to be a $\mathbb{Z}_N$ two-form gauge field.
The gauged theory is essentially the Maxwell theory at a coupling ${\tau}/{N^2}$ and the background gauge field $B$ couples to the $\mathbb{Z}_N^{(1)}$ subgroup of the $U(1)^{(1)}$ magnetic one-form symmetry. In particular, since we start with a theory at $\theta=\pi pN$, the gauged theory is at $\theta=\pi p/N$.

The important point is that the gauged theory  is invariant under the $CTST$ gauging of its $\mathbb{Z}_N^{(1)}$ magnetic one-form symmetry followed by a time-reversal transformation $K$. 
This follows  from the anomalous time-reversal transformation \eqref{eq:anomaly_TR}:
\ie
&\,\mathcal{L}_{\mathcal{Q}_{\tau/N^2}}[B,0]
\\
\xmapsto{CTST}&\, \mathcal{L}_{\mathcal{Q}_\tau}[0,c]-\frac{pN}{4\pi}c\wedge c+\frac{N}{2\pi} c\wedge du+\frac{N}{2\pi}c\wedge B \\
\xmapsto{\ \ \ K\ \ \ }&\,\mathcal{L}_{\mathcal{Q}_\tau}[0,c]+\frac{N}{2\pi} c\wedge du+\frac{N}{2\pi}c\wedge B
= \mathcal{L}_{\mathcal{Q}_{\tau/N^2}}[B,0]\,.
\fe
This leads to a non-invertible time-reversal symmetry $\mathsf{T}^{\theta=\pi p/N}$ associated with the $KCTST$ transformation at $\theta=\pi p/N$, which is the one discussed in Sections \ref{sec:maxwellnoninv} and \ref{sec:gauging}. 

For even $N$, the non-invertible time-reversal symmetry $\mathsf{T}^{\theta=\pi p/N}$ can be interpreted as a consequence of the mixed anomaly between the time-reversal symmetry $\mathsf{T}^{\theta=\pi pN}$ and the $\mathbb{Z}_N^{(1)}$ electric one-form symmetry at $\theta=\pi pN$.  
This construction for the non-invertible symmetry from an anomaly via gauging falls into the same paradigm as in \cite{Tachikawa:2017gyf,Thorngren:2018bhj,Ji:2019ugf,Kaidi:2021xfk}.

However,  for odd $N$, the anomaly is trivial at $\theta=\pi pN$ so the same interpretation does not hold. Then, why is there still a non-invertible time-reversal symmetry at $\theta=\pi p/N$ for odd $N$? This is because in order to trivialize the anomaly at $\theta=\pi pN$, we need to choose a particular counterterm for the background gauge field $C$ as in \eqref{eq:counterterm}.  
On the other hand, when we gauge the symmetry to obtain ${\cal Q}_{\tau/N^2}$, we choose a different counterterm. 
Hence, the anomalous transformation is not trivialized and  leads to the non-invertible time-reversal symmetry $\mathsf{T}^{\theta=\pi p/N}$ in the gauged theory.  
This is a new mechanism for a non-invertible symmetry to arise from gauging, without having an anomaly in the first place. 
In Appendix \ref{app:PSUN}, we find similar non-invertible symmetries from this new mechanism in the $PSU(N)$ Yang-Mills theory at $\theta\in \pi\mathbb{Z}$ for odd $N$, generalizing the construction of \cite{Kaidi:2021xfk}.

\section{Massive QED}

We will assume that there is no dynamical monopole in the scale we are interested in. 
Under this assumption, the massive QED has a magnetic $U(1)^{(1)}$ one-form symmetry, while the electric $U(1)^{(1)}$ one-form symmetry of the free Maxwell theory is broken by the electron. 
Using the magnetic symmetry, the  same construction of the non-invertible time-reversal symmetry in the free Maxwell theory without matter in Sections \ref{sec:maxwellnoninv} and \ref{sec:gauging} applies to  massive QED.  

\begin{figure}[!t]
    \centering
    \includegraphics[width=0.6\textwidth]{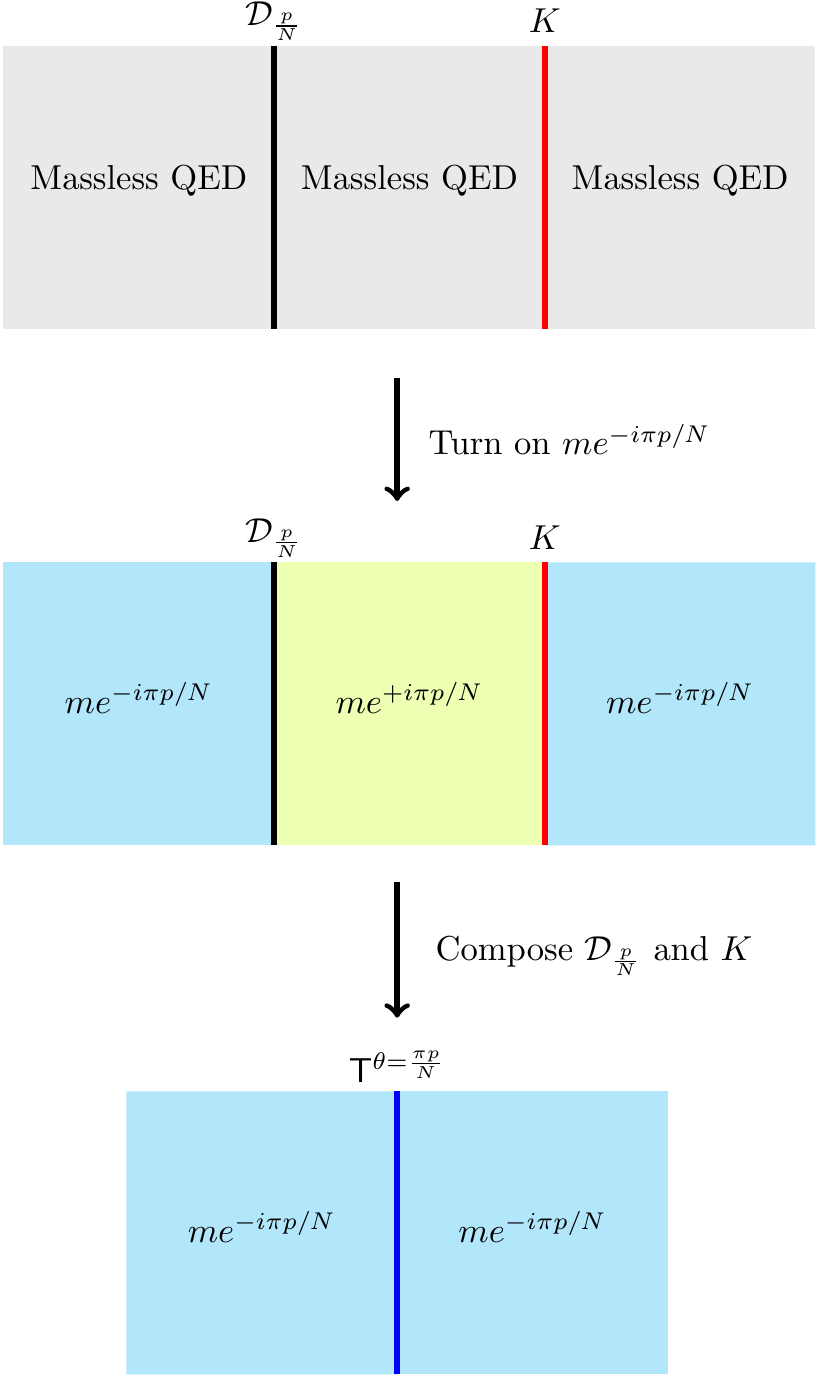}
    \caption{The massless QED has an infinite linear non-invertible symmetry generated by  $\mathcal{D}_\frac{p}{N}$ \cite{Choi:2022jqy,Cordova:2022ieu}, and an invertible time-reversal symmetry generated by $K$.
    Once we turn on a complex mass term $me^{-i\pi p/N}$, both of these symmetries are explicitly broken.
    The defect $\mathcal{D}_\frac{p}{N}$ now becomes a topological interface separating two massive QED theories with different mass parameters $me^{-i\pi p/N}$ and $me^{+i\pi p/N}$, and similarly for $K$.
    The composition of these two symmetries is a non-invertible time-reversal symmetry operator $\mathsf{T}^{\theta=\frac{\pi p}{N}}$ preserved by the massive QED at $\theta= \pi p/N$.
    }
    \label{Fig:QED_interface}
\end{figure}
  
Consider  QED with a single Dirac fermion with mass $m$.
The Lagrangian is given by
\begin{equation}
    -\frac{1}{4e^2} F_{\mu\nu}F^{\mu\nu} + i\bar{\Psi} (\partial_\mu - i A_\mu) \gamma^\mu  \Psi
    + m \bar{\Psi} \Psi
    + \frac{\theta}{32\pi^2} \varepsilon_{\mu\nu\rho\sigma} 
    F^{\mu\nu} F^{\rho\sigma}  \,. 
\end{equation}
Using a chiral rotation, we  can take $m>0$ to be a positive real constant, so that the $\theta$-angle is physical.  
Alternatively, we can set the $\theta$-angle to zero and work with a complex mass term $m e^{-i\theta}$ whose phase is physical.

At $\theta=0$ and $\theta=\pi$, massive QED has an invertible time-reversal symmetry $K$ and $\mathbb{T}K$, respectively. 
The choice of $K$ is not unique: one can compose $K$ with a unitary $\mathbb{Z}_2^{(0)}$ symmetry to obtain another time-reversal symmetry. 
We will not commit to any particular choice of $K$.

At $\theta = \pi p/N$, the theory is invariant under the $KCTS T$ transformation.
Thus, the massive QED also has the same kind of a non-invertible time-reversal symmetry $\mathsf{T}^{\theta=\frac{\pi p}{N}} = K \circ \mathsf{I}_{2\pi p \over N}$ as in the Maxwell theory.  
At low energy, we can integrate out the fermions and the massive QED flows to the free Maxwell theory with the same value of the $\theta$-angle.
The non-invertible time-reversal symmetry is matched along this renormalization group flow.

\begin{figure}[!t]
    \centering
    \includegraphics[width=0.7\textwidth]{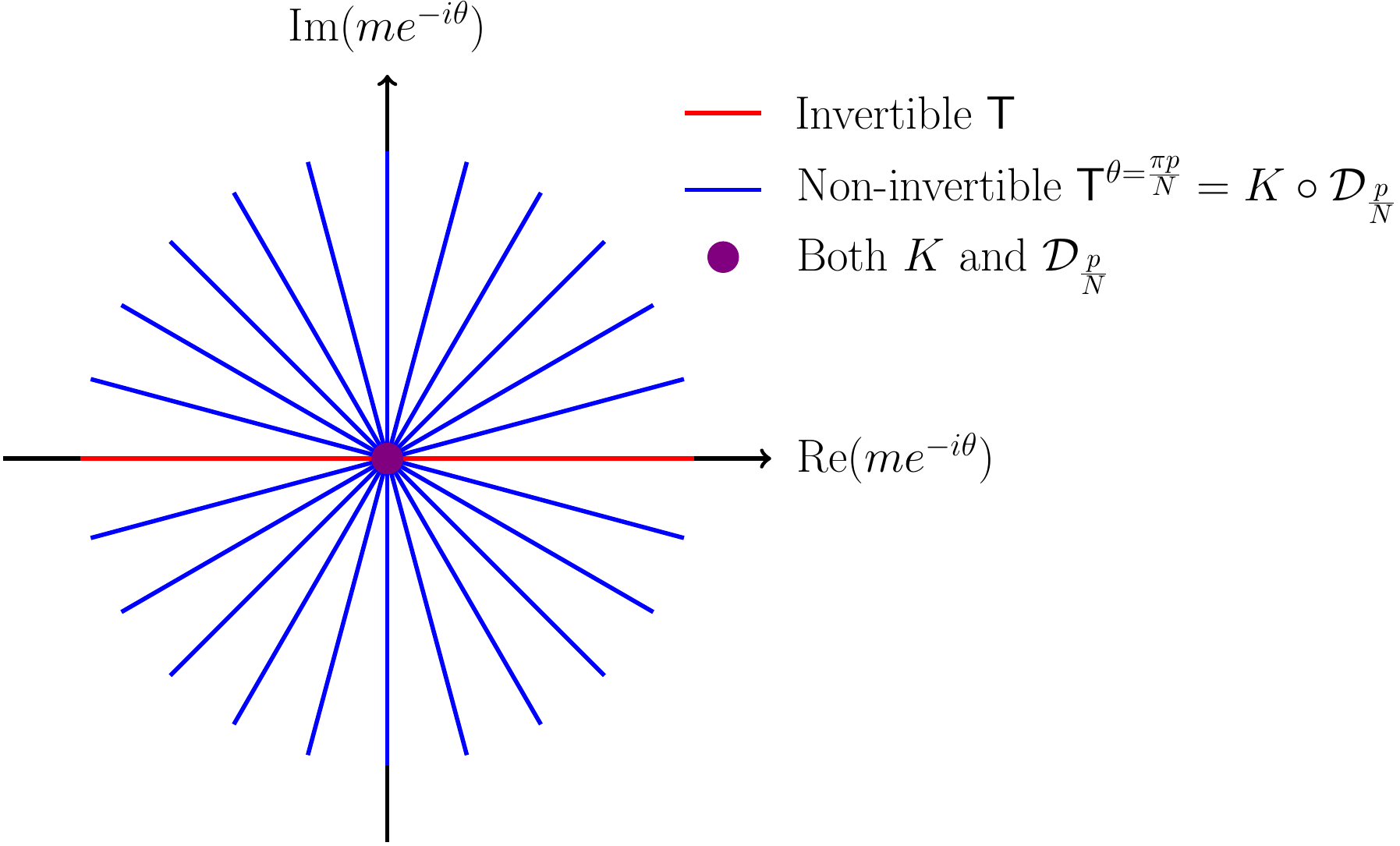}
    \caption{Invertible and non-invertible time-reversal symmetries of QED on the complex mass plane.
    If the mass $m e^{-i\theta}$ is real (i.e., if $\theta=0,\pi$), the theory has an invertible time-reversal symmetry.
    When the phase of the mass term is rational, i.e., $\theta=\pi p/N$, we have the non-invertible time-reversal symmetry generated by $\mathsf{T}^{\theta=\frac{\pi p}{N}}$.
    At zero mass, we have the infinite linear non-invertible symmetry generated by $\mathcal{D}_{\frac{p}{N}}$ for all rational numbers $p/N$ and the invertible time-reversal symmetry.
    } 
    \label{Fig:QED_phase_diagram}
\end{figure}

There is an alternative construction for $\mathsf{T}^{\theta=\frac{\pi p}{N}}$ in massive QED from the linear, non-invertible symmetry in massless QED.  
It was realized in \cite{Choi:2022jqy,Cordova:2022ieu} that the classical axial $U(1)_A$ symmetry of massless QED is not completely broken by the ABJ anomaly, but it turns into a linear, non-invertible symmetry ${\cal D}_{p\over N}$ labeled by the rational numbers $p/N$.  
In addition,  the massless QED has an invertible time-reversal symmetry $K$. 
(When $m=0$,  the $\theta$-angle is not physical and  can be set to zero by an axial rotation.)

Now we turn on a mass term $me^{-i\theta}$, and the massless QED flows to the massive QED with a $\theta$-angle. 
For a generic value of $\theta$, the mass deformation breaks both the invertible time-reversal symmetry  $K$ and the non-invertible symmetry ${\cal D}_{p\over N}$ (which is associated with the $CTST$ transformation). 
At a rational value of $\theta=\pi p/N$, the composition of $K$ and ${\cal D}_{p\over N}$  is  preserved by the mass deformation $me^{-i\pi p/N}$.  
The composed operator  implements a non-invertible time-reversal symmetry 
\ie
\mathsf{T}^{\theta= {\pi p\over N}} = K \circ {\cal D}_{p\over N} \,.
\fe
See  Figure \ref{Fig:QED_interface}. 
In particular, if we set $p=N=1$, then $\mathcal{D}_{\frac{p}{N}}$ becomes an invertible defect shifting the $\theta$-angle by $-2\pi$, that is, it reduces to the $\mathbb{T}^{-1}$ transformation.
Correspondingly, the invertible time-reversal symmetry at $\theta=\pi$ is associated with the $K\mathbb{T}^{-1} = \mathbb{T}K$ transformation.

Using ${\cal D}_{p\over N}\times {\cal D}_{p\over N}^\dagger={\cal D}_{p\over N}^\dagger\times {\cal D}_{p\over N}= {\cal C}^{(N)}$ from \cite{Choi:2022jqy}, we immediately recover the  non-invertible fusion rule \eqref{fusion}. 
In Figure \ref{Fig:QED_phase_diagram}, we summarize the invertible and non-invertible symmetries of massive QED across the complex mass plane.

\section{$\mathcal{N}=4$ Super Yang-Mills Theory}\label{sec:N=4}

In this section we point out similar non-invertible time-reversal symmetries in the 3+1d ${\cal N}=4$ $SU(2)$ super Yang-Mills theory. 
We  then relate these non-invertible time-reversal symmetries to various linear and anti-linear non-invertible symmetries discovered in \cite{Choi:2021kmx,Kaidi:2021xfk,Choi:2022zal,Kaidi:2022uux}. 
Our discussion  makes use of many results  reported in  \cite{Kaidi:2022uux}. 
For simplicity, we will assume the spacetime manifold to be spin and in particular oriented.

The ${\cal N}=4$ $SU(2)$ super Yang-Mills theory is parametrized by  a complexified coupling $\tau$. 
Similar to the free Maxwell theory, it enjoys an $SL(2,\mathbb{Z})$ electromagnetic duality, but the details differ as we discuss below. 
While the $\mathbb{T}: \tau\mapsto \tau+1$ transformation is an exact duality of the $SU(2)$ theory, the $\mathbb{S}:\tau\mapsto -1/\tau$ transformation maps the latter to a theory with an $SO(3)$ gauge group. 
The $SL(2,\mathbb{Z})$ duality transformation  acts non-trivially on the spectrum of line operators \cite{Kapustin:2005py,Gaiotto:2010be,Aharony:2013hda}. 
The $SU(2)$ theory has a $\mathbb{Z}_2^{(1)}$ center one-form symmetry, for which we can define a projective $SL(2,\mathbb{Z}_2)$ transformation generated by  the $S,T,C$  in \eqref{eq:sigma_tau_operations}  via gauging.

Let  us start with  the anti-linear, invertible symmetries of the $SU(2)$ theory. 
Along $\theta=0$, there is an invertible time-reversal symmetry $\mathsf{T}^{\theta=0}$ associated with the naive time-reversal transform $K$.\footnote{Again, one can always compose one time-reversal  transformation with a unitary $\mathbb{Z}_2^{(0)}$ symmetry to obtain another time-reversal transformation. We will not commit to a specific choice of $K$ at $\theta=0$. }
Along $\theta=\pi$, the coupling constant $\tau$ is invariant under the $\mathbb{T} K$ transformation. If we further track the dependence on the coupling to the background gauge field of the $\mathbb{Z}_2^{(1)}$ one-form symmetry, then the $SU(2)$ theory at $\theta=\pi$ is invariant under $T\mathbb{T}K= \mathbb{T}TK$   \cite{Kaidi:2022uux}. We denote the corresponding invertible time-reversal symmetry by $\mathsf{T}^{\theta=\pi}$.
Similarly, at $\theta = -\pi$, we have an invertible time-reversal symmetry $\mathsf{T}^{\theta=-\pi}$ that is associated with the transformation $\mathbb{T}^{-1} TK$.

Next, we review the linear, non-invertible symmetries of the $SU(2)$ theory. 
It was shown in \cite{Kaidi:2021xfk,Choi:2022zal,Kaidi:2022uux} that at $\tau=i$, there is a  duality defect ${\cal D}_2$ associated with the transformation $\mathbb{S}S$.  
Importantly, while this duality defect acts invertibly on the local operators as a $\mathbb{Z}_4^{(0)}$ symmetry, it acts non-invertibly on the line operators. 
In other words, the $\mathbb{S}$-duality defect at $\tau=i$ is  non-invertible. 
 Similarly, at $\tau=e^{2\pi i /3}$, there is a triality defect ${\cal D}_3$ associated with the transformation $\mathbb{ST}ST$  \cite{Choi:2022zal,Kaidi:2022uux}. 
At these two special points, we can compose the duality and triality defects with the invertible time-reversal symmetries to obtain non-invertible time-reversal symmetries, i.e.,  ${\cal D}_2\circ\mathsf{T}^{\theta=0}$ at $\tau=i$ and  ${\cal D}_3\circ \mathsf{T}^{\theta=-\pi}$ at $\tau=e^{2\pi i /3}$.

We now move on to the more general non-invertible time-reversal symmetry along the locus $|\tau|=1$.  
While the transformation $\mathbb{S}K$ leaves $\tau$ invariant, unlike the Maxwell theory, it maps the $SU(2)$ theory to an $SO(3)$ theory. 
To remedy this, we compose the above transformation with the $S$ transformation, which gauges the $\mathbb{Z}_2^{(1)}$ one-form symmetry. The composite transformation $\mathbb{S}S K$  then leaves the $SU(2)$ theory at any point along $|\tau|=1$ invariant.  
We denote the corresponding anti-linear symmetry by $\mathsf{T}^{|\tau|=1}$. 
Since $\mathsf{T}^{|\tau|=1}$ involves the $S$ transformation, it is non-invertible. We have
\ie
(\mathsf{T}^{|\tau|=1} )^\dagger \times \mathsf{T}^{|\tau|=1}
= {\cal C}^{(2)}_0 \equiv {1\over |H^0(M,\mathbb{Z}_2)|} \sum_{\Sigma\in H_2(M,\mathbb{Z}_2)} \eta(\Sigma)\,,
\fe
where $\eta(\Sigma)$ is the $\mathbb{Z}_2^{(1)}$ one-form symmetry operator on a two-surface $\Sigma$, and $M$ is the three-manifold on which $\mathsf{T}^{|\tau|=1}$ is supported.  ${\cal C}^{(2)}_0$ is the untwisted condensation defect from one-gauging the $\mathbb{Z}_2^{(1)}$ one-form center symmetry on $M$ (see Appendix \ref{app:condensation} for more details).

At  $\tau=i$, the non-invertible time-reversal symmetry $\mathsf{T}^{|\tau|=1}$ is a composition of the duality defect ${\cal D}_2$ (which is associated with $\mathbb{S}S$) and the invertible time-reversal symmetry $\mathsf{T}^{\theta=0}$ (which is associated with $K$). Similarly, at $\tau=e^{2\pi i/3}$, $\mathsf{T}^{|\tau|=1}$ is the composition of the triality defect ${\cal D}_3$ (which is associated with $\mathbb{S}\mathbb{T}ST$) and   the invertible time-reversal symmetry $\mathsf{T}^{\theta=-\pi}$ (which is associated with $\mathbb{T}^{-1} T K$).

Finally, we note that the locus $|\tau|=1$ of the $SU(2)$ theory is mapped to $\theta=\pi$ of the $SO(3)_-$ theory under the $\mathbb{TST}$ duality transformation (up to a counterterm of $\mathbb{Z}_2^{(1)}$ which can be fixed by a $T$ transformation). 
In \cite{Kaidi:2021xfk}, it was found that the $SO(3)_-$ pure Yang-Mills theory has  a non-invertible time-reversal symmetry at $\theta=\pi$. (Recall that the $\theta$-angle is $4\pi$-periodic for the $SO(3)$ gauge group.)  
When embedded into the ${\cal N}=4$ $SO(3)_-$ theory, this non-invertible time-reversal symmetry is related to our $\mathsf{T}^{|\tau|=1}$ of the $SU(2)$ theory by an electromagnetic duality transformation.

Intuitively, the $\mathbb{Z}_2^{(1)}$ one-form symmetry of the $SO(3)_-$ theory makes it possible to define a non-invertible time-reversal symmetry at half of the naive allowed values of $\theta$, i.e., $\theta\in 2\pi \mathbb{Z}$. 
In contrast, the $U(1)^{(1)}$ magnetic one-form symmetry cures the time-reversal symmetry  at every rational $\theta$-angle for the abelian gauge theory. 

 \begin{table} 
    \centering
    \begin{tabular}{|c|c|c|c|} 
        \hline
~complexified coupling ~$\tau$~& ~symmetry~&~~ invertible?  ~~&~~linear?~~\\
      \hline
        $\theta=0$ & $K$   & Yes&anti-linear\\
        \hline
                $\theta=\pi $ & $\mathbb{T}T K$   & Yes&~~anti-linear~~\\
                \hline
                 $|\tau|=1$ & $\mathbb{S}S K$   & No& anti-linear\\
\hline
$\tau= i $  &  ~~$\mathbb{S}S$~~  & No&linear\\
\hline
$\tau= e^{2\pi i/3} $  &  ~~$\mathbb{ST}ST$~~  & No&linear\\
        \hline
    \end{tabular}
    \caption{Some (anti-)linear (non-)invertible symmetries of the ${\cal N}=4$ $SU(2)$ super Yang-Mills theory. Here $K$ is the naive time-reversal transformation, $\mathbb{S,T}$ generate the electromagnetic duality, and $S,T$ generate the (projective) $SL(2,\mathbb{Z}_2)$ transformation via gauging the $\mathbb{Z}_2^{(1)}$ center one-form symmetry. 
    }
    \label{table:N=4}
\end{table}

\section{Trivially Gapped Phases} \label{sec:SPT_analysis}

We have discussed non-invertible time-reversal symmetries in the free Maxwell theory, massive QED, and ${\cal N}=4$ super Yang-Mills theory. 
These symmetries  always arise from the  invariance of a QFT under the $CKS = SK$ transformation  (provided we choose the appropriate counterterm of the background gauge field for the $\mathbb{Z}_N^{(1)}$ one-form symmetry, and up to a duality transformation in the case of the $\mathcal{N}=4$ super Yang-Mills theory).
We now study these symmetries in a trivially gapped phase with a $\mathbb{Z}_N^{(1)}$ one-form global symmetry i.e. a $\mathbb{Z}_N^{(1)}$ one-form SPT phase. 
We will consider oriented SPT phases, both bosonic and fermionic.\footnote{For  discussions about time-reversal symmetric SPT phases with a fusion category symmetry in 1+1 dimensions, see \cite{Inamura:2021wuo}. }

This SPT analysis is relevant for the generalized anomalies of these non-invertible time-reversal symmetries, and our discussion in this section is an anti-linear version of the one in \cite{Choi:2021kmx}. 
Starting from  a QFT with a non-invertible time-reversal symmetry, we ask whether there can be a renormalization group flow to a trivially gapped phase while preserving this symmetry. 
If not, this signals a generalized 't Hooft anomaly for the symmetry.
If yes, this trivially gapped phase must be described by  a $\mathbb{Z}_N^{(1)}$-SPT phase that is invariant under the $CKS$ transformation.

We emphasize that finding an SPT phase that is invariant under the topological transformation (e.g., discrete gauging) is only  a necessary condition for the  associated generalized symmetry to be anomaly-free (in the sense that it is compatible with a trivially gapped phase), but not sufficient. 
    For instance, consider a $\mathbb{Z}_2^{(0)} \times \mathbb{Z}_2^{(0)}$ zero-form symmetry in 1+1-dimensions. 
      In general, if a 1+1d QFT is invariant under gauging the $\mathbb{Z}_2^{(0)} \times \mathbb{Z}_2^{(0)}$ symmetry, then it realizes a $\mathbb{Z}_2^{(0)} \times \mathbb{Z}_2^{(0)}$ Tambara-Yamagami fusion category symmetry \cite{TAMBARA1998692}. 
    It is easy to check that there exists a $\mathbb{Z}_2^{(0)} \times \mathbb{Z}_2^{(0)}$-SPT phase which is invariant under gauging the $\mathbb{Z}_2^{(0)} \times \mathbb{Z}_2^{(0)}$ symmetry.
    However, such a QFT may or may not be able to be trivially gapped in the IR while preserving the symmetries, depending on whether the corresponding fusion category admits a fiber functor or not.
    There are four inequivalent $\mathbb{Z}_2^{(0)} \times \mathbb{Z}_2^{(0)}$ Tambara-Yamagami fusion categories, among which only three of them admit a fiber functor \cite{Bhardwaj:2017xup,Thorngren:2019iar}.
    Thus, even though there exists a $\mathbb{Z}_2^{(0)} \times \mathbb{Z}_2^{(0)}$-SPT invariant under gauging the $\mathbb{Z}_2^{(0)} \times \mathbb{Z}_2^{(0)}$ symmetry, there might be additional obstructions to trivially gap a QFT with a $\mathbb{Z}_2^{(0)} \times \mathbb{Z}_2^{(0)}$ Tambara-Yamagami fusion category symmetry.

\subsection{Odd $N$}

For odd $N$, the possible SPT phases are labeled by an integer $q \sim q+N$, where the partition function is given by\footnote{Here, the normalization of the $\mathbb{Z}_N^{(1)}$ background gauge field $B$ is different from the previous sections. In this section, $B$ denotes the discrete $\mathbb{Z}_N$ two-cocycle, and its values can be any integer modulo $N$. The same convention applies for the dynamical gauge field $b$.}
\begin{equation} \label{eq:spt_odd_N}
    \exp \left( \frac{2\pi i q}{N} \int B \cup B \right) \,.
\end{equation}
If we apply the $S$ transformation, the resulting theory is again invertible if and only if $\text{gcd}(q,N) = 1$.
Under this condition, we obtain
\begin{equation}
    \sum_b \exp\left[
      \int  \left( \frac{2\pi i q}{N} b \cup b +
      \frac{2\pi i k}{N} b \cup B
      \right)
    \right] 
    = \exp \left[\frac{2\pi i }{N} \left(
        -\frac{k^2}{4q}
    \right)\int B \cup B \right] \,,
\end{equation}
where we are neglecting the overall normalization and the gravitational counterterms.
Thus, we see that under the $S$ transformation, the SPT phase transforms as $q \mapsto -k^2/4q$.

On the other hand, the $K$ transformation simply acts as complex conjugation on the partition function.
Therefore, under the $KS$ transformation, we have
\begin{equation}
    q \xmapsto{S} -k^2/4q \xmapsto{K} +k^2/4q \,.
\end{equation}
The $C$ transformation acts trivially on the SPT phase \eqref{eq:spt_odd_N}, since its action is quadratic in $B$.
Thus, the $CKS$ invariance of the SPT phase requires $q = +k^2/4q$ mod $N$, or equivalently
\begin{equation}
    4q^2 = k^2 \quad \text{mod $N$.} 
\end{equation}
Since $N$ is odd, 2 is invertible mod $N$, and $q= 2^{-1} k$ mod $N$ is always a solution of the equation.
We conclude that the $CKS$ invariance is always compatible with a trivially gapped phase for any $N$ and $p$.

\subsection{Even $N$}

For even $N$, the possible SPT phases are
\begin{equation} \label{eq:spt_even_N}
   \exp \left( \frac{2\pi i q}{2N} \int \mathcal{P}(B) \right) \,,
\end{equation}
where $\mathcal{P}(B)$ is the Pontryagin square of $B$.
We have $q \sim q+2N$ for the bosonic case, and $q \sim q+N$ for the fermionic case.
In the even $N$ case, the $S$ transformation acts as
\begin{equation}
    \sum_b \exp\left[
      \int  \left( \frac{2\pi i q}{2N} \mathcal{P}(b) +
      \frac{2\pi i k}{N} b \cup B
      \right)
    \right] 
    = \exp \left[\frac{2\pi i }{2N} \left(
        -\frac{k^2}{q}
    \right)\int \mathcal{P}(B) \right] \,.
\end{equation}
That is, we get $q \mapsto -k^2/q$ under $S$.

Combined with the $K$ transformation, we obtain
\begin{equation}
    q \xmapsto{S} -k^2/q \xmapsto{K} +k^2/q \,.
\end{equation}
Again, $C$ acts trivially on the SPT phases \eqref{eq:spt_even_N}.
Therefore, the condition for the SPT phase to be $CKS$ invariant is 
\begin{equation}
    q^2 =k^2 \quad
    \begin{cases}
    \text{mod~ $2N$~  if bosonic}\\
        \text{mod~  $N$~  if fermionic} 
    \end{cases}\,.
\end{equation}
We see that $q=k$ is always a solution, thus again the invariance under the $CKS$ transformation is always compatible with a trivially gapped phase.

Since in each of the above cases we find an SPT phase that is invariant under the $CKS$ transformation, we cannot conclude if there are obstructions to realizing the non-invertible time-reversal symmetry in a trivially gapped phase. 
This is to be contrasted with the analyses in \cite{Choi:2021kmx,Choi:2022zal} where certain linear, non-invertible symmetries are found to be intrinsically incompatible with a trivially gapped phase.

\section{Conclusion and Outlook}

We find that the free Maxwell theory and massive QED for the real world at a rational $\theta$-angle is time-reversal invariant. 
The time-reversal symmetry is implemented by a conserved, anti-linear operator $\mathsf{T}^{\theta=\pi p/N}$, but it is  non-invertible because of the fractional quantum Hall state attached to the operator.

In this work, we only identify the existence of these new time-reversal symmetries, and defer their applications for future investigations. Below we outline a few interesting future directions:

\begin{itemize}

\item In the context of the strong $\mathsf{CP}$ problem, it is natural to wonder if there is a hidden non-invertible $\mathsf{T}$ or $\mathsf{CP}$  symmetry when the $SU(3)$ $\theta$-angle vanishes in the Standard Model.\footnote{In the Standard Model, there is a combination of the $SU(2)\times U(1)_Y$ $\theta$-angles that is physical and cannot be removed by chiral rotations \cite{Tong:2017oea}. 
This physical $\theta$-angle is another source of $\mathsf{CP}$ violation, but it is rarely discussed in the strong $\mathsf{CP}$ problem. Unlike its $SU(3)$ counterpart, this $\theta$-angle does not affect  any known experimental observables such as the neutron electric dipole moment.}

\item   A time-reversal invariant quantum field theory can be placed on an unoriented manifold by inserting the time-reversal defect along the orientation-reversal cut of the manifold \cite{Witten:2016cio}.\footnote{When the time-reversal symmetry is anomalous,  the partition function on an unoriented manifold might be subject to a phase ambiguity. See, for example, \cite{Tachikawa:2016cha,Tachikawa:2016nmo}. } Since the non-invertible time-reversal symmetry operators contain a fractional quantum Hall state, one needs to place these topological degrees of freedom along the cut as well.\footnote{We thank K.\ Ohmori for discussions on this point.}  More generally, how do we describe the  tangential structure associated with the non-invertible time-reversal symmetry? 

How do we characterize the generalized anomalies of these non-invertible time-reversal symmetries?

\item
It is known that at $\theta=0$, there are 7 different versions of time-reversal symmetric Maxwell theory, distinguished by the quantum numbers of the Wilson and 't Hooft lines \cite{Wang:2015fmi,Hsin:2019fhf}.\footnote{We thank S.\ Seifnashri for discussions on related points.} It would be interesting to generalize this discussion to Maxwell theory with non-invertible time-reversal symmetry at a rational $\theta$-angle.

\item Related to the previous point,  it would be interesting to compute the partition function of the Maxwell theory at $\theta=\pi p/N$ on an unoriented manifold using our non-invertible time-reversal symmetry, extending the classic results of \cite{Witten:1995gf,Metlitski:2015yqa}.  Similar generalizations of \cite{Wang:2020jgh,Caetano:2022mus} can be explored  for the ${\cal N}=4$ super Yang-Mills theory along $|\tau|=1$.

\item 
By treating the gauge field $A$ in Section \ref{sec:maxwellnoninv} as a classical gauge field, we find that the invertible phase with the classical action $\frac{\theta}{8\pi^2} \int F\wedge F$ at $\theta=\frac{\pi p}{N}$ also has a non-invertible time-reversal symmetry \eqref{eq:noninv_time_reversal}. This can be viewed as an SPT phase protected by a $U(1)^{(0)}$ symmetry and a non-invertible time-reversal symmetry, generalizing the usual 3+1-dimensional topological insulator at $\theta=\pi$ with an invertible time-reversal symmetry.\footnote{This is to be contrasted with a 3+1-dimensional fractional topological insulator, whose electromagnetic response is also captured by the same effective action $\frac{\theta}{8\pi^2} \int F \wedge F$ with a fractional value of $\theta$. See, for instance, \cite{Maciejko:2010tx,Maciejko:2011ed,PhysRevB.83.195139,PhysRevB.94.115104,Ye:2017axd}.
A fractional topological insulator at low energy is not an SPT phase but a nontrivial TQFT with a $U(1)^{(0)}$ and an invertible time-reversal symmetry, i.e., a symmetry enriched topological (SET) phase. It is described by the following action at $\theta=\frac{\pi p}{q^2}$ \cite{Ye:2017axd},
\ie\label{eq:action_FTI}
\frac{q}{2\pi}\int  b\wedge da+\frac{1}{2\pi}\int b\wedge dA + \frac{\theta}{8\pi^2}\int F\wedge F~,
\fe
where $b$ and $a$ are dynamical two-form and one-form gauge field, respectively for the low-energy $\mathbb{Z}_q$ gauge theory. Interestingly, in addition to the above values of the $\theta$-angle, the action \eqref{eq:action_FTI} at  $\theta=\frac{\pi p}{q^2 N}$ is invariant under a non-invertible time-reversal symmetry. Similar to the construction at Section \ref{sec:gauging}, this non-invertible time-reversal symmetry arises from the invariance under a $CTST$ gauging of a $\mathbb{Z}_N^{(1)}$ one-form symmetry (which couples to the theory as $\frac{k}{2\pi}\int B\wedge da$), followed by a naive time-reversal transformation $K$. To conclude,  \eqref{eq:action_FTI} at $\theta=\frac{\pi p}{q^2N}$ is a generalized fractional topological insulator with a non-invertible time-reversal symmetry.
} 

\end{itemize}

\section*{Acknowledgements}

We are grateful to C.\ Cordova, J.\ Kaidi, K.\ Ohmori, N.\ Seiberg, and S.\ Seifnashri  for useful discussions. 
HTL is supported in part by a Croucher fellowship from the Croucher Foundation, the Packard Foundation and the Center for Theoretical Physics at MIT. 
The work of SHS was supported in part by NSF grant PHY-2210182. 
We thank the Simons Collaboration on Global Categorical Symmetries and the Perimeter Institute for Theoretical Physics for their hospitality during a conference and a school.
Research at Perimeter Institute is supported in part by the Government of Canada through the Department of Innovation, Science and Economic Development Canada and by the Province of Ontario through the Ministry of Colleges and Universities. 
We also thank the Simons Summer Workshop for hospitality during the final stage of this project. 
The authors of this paper were ordered alphabetically.

\appendix

\section{Condensation Defect}\label{app:condensation}

In this appendix, we review the construction of the condensation defects/operators in the Maxwell theory that arise from one-gauging \cite{Roumpedakis:2022aik} a $\mathbb{Z}_N^{(1)}$ subgroup of the magnetic one-form symmetry along a three-dimensional spin submanifold $M$ in  spacetime. 
See \cite{Choi:2022zal,Choi:2022jqy} for more details.

Such a condensation defect in general takes the  form:
\begin{equation} \label{eq:cond_defect_1}
    \mathcal{C}^{(N)}_\epsilon = \frac{1}{|H^0(M,\mathbb{Z}_N)|}
    \sum_{\Sigma \in H_2 (M,\mathbb{Z}_N)} \epsilon(M,\Sigma) \eta(\Sigma) \,.
\end{equation}
Here, $\eta(\Sigma) = \exp \left( \frac{i}{N} \oint_\Sigma dA \right)$ is the $\mathbb{Z}_N^{(1)}$ magnetic one-form symmetry operator supported on a two-cycle $\Sigma$, and $\epsilon(M,\Sigma) \in \text{Hom}(\text{Tors} \,\Omega_3^{\text{Spin}}(B\mathbb{Z}_N),U(1))$ is a discrete torsion phase.
 
Different choices of the discrete torsion lead to different condensation operators. 
In \cite{Choi:2022jqy}, it was shown that a subset of these condensation operators in the Maxwell theory can be described in terms of the worldvolume Lagrangian
\begin{equation} \label{eq:cond_defect_2}
    \mathcal{C}_K^{(N)} = \exp \left[
        i \oint_M \left(
            \frac{N}{2\pi} a \wedge d\tilde{a} + \frac{K}{4\pi} a \wedge da + \frac{1}{2\pi} a \wedge dA
        \right)
    \right] \,,
\end{equation}
where $K$ is an integer specifying the discrete torsion, and $a$ and $\tilde{a}$ are dynamical $U(1)$ one-form gauge fields that live only on $M$.
Here $\tilde{a}$ serves as a Lagrange multiplier. 
Once we integrate out $\tilde{a}$, the gauge field $a$ is forced to be a $\mathbb{Z}_N$ one-form gauge field living on $M$.
The remaining path integral over $a$ collapses into a finite sum over $H^1(M,\mathbb{Z}_N)$, which is isomorphic to $H_2(M,\mathbb{Z}_N)$ via the Poincar\'e duality.
The third term in the worldvolume Lagrangian \eqref{eq:cond_defect_2} reduces to $\exp (i\oint_M \frac{1}{2\pi} a \wedge dA) = \exp (\frac{i}{N}\oint_\Sigma dA) = \eta(\Sigma)$, where $\Sigma \in H_2(M,\mathbb{Z}_N)$ is the Poincar\'e dual of the gauge field $a$ on $M$. 
The expression \eqref{eq:cond_defect_2} then reduces to the usual definition of the condensation operator given in \eqref{eq:cond_defect_1}.
The middle term $\frac{K}{4\pi} a \wedge da$ in the worldvolume Lagrangian corresponds to a particular choice of the discrete torsion $\epsilon(M,\Sigma)$. 

As was explained in \cite{Choi:2022jqy}, the worldvolume Lagrangian \eqref{eq:cond_defect_2} corresponds to the 2+1d $\mathbb{Z}_N$ gauge theory at level $K$ living on $M$, where the background gauge field for its $\mathbb{Z}_N^{(1)}$ one-form symmetry is activated using the 3+1d bulk field strength $dA/N$, properly normalized.
As such, the integer $K$ has a finite periodicity.
For odd $N$, condensation operators labeled by $K$ and $K+N$ are equivalent, that is, $K \sim K+N$.
For even $N$, the periodicity of $K$ is $2N$, that is, $K \sim K+2N$.

Among the condensation operators \eqref{eq:cond_defect_2}, the one with $K=N$ (which is equivalent to $K=0$ for odd $N$) appears in the fusion algebra of the non-invertible time-reversal symmetries.
We define
\begin{equation}
    \mathcal{C}^{(N)} \equiv \begin{cases}
        \mathcal{C}^{(N)}_N &\quad \text{if $N$ is even,} \\
        \mathcal{C}^{(N)}_0 &\quad \text{if $N$ is odd.}
    \end{cases}
\end{equation}
Indeed, in  \eqref{eq:fusion}, we obtained
\begin{equation}
    (\mathsf{T}^{\theta={ \pi \over N}} )^\dagger\times \mathsf{T}^{\theta={ \pi \over N}} 
= \exp \left[        i\oint_M \left(    \frac{N}{4\pi} ada  - {N\over 4\pi} \bar ad\bar a +\frac{1}{2\pi}( a-\bar a)dA
        \right)
    \right] \,.
\end{equation}
On the right-hand side, if we perform a field redefinition $a \rightarrow a+\bar{a}$, we see that the expression reduces to the definition of $\mathcal{C}^{(N)}$.
More generally, we have $(\mathsf{T}^{\theta={ \pi p \over N}} )^\dagger\times \mathsf{T}^{\theta={ \pi p \over N}} = \mathcal{C}^{(N)}$ as claimed.

\section{$PSU(N)$ Yang-Mills Theory at $\theta\in\pi \mathbb{Z}$}\label{app:PSUN}

It was pointed out in \cite{Kaidi:2021xfk} that there is a non-invertible time-reversal symmetry in the $PSU(N)$ Yang-Mills theory at $\theta=\pi$ for even $N$. The non-invertible symmetry arises from gauging the electric $\mathbb{Z}_N^{(1)}$ one-form symmetry, which has a mixed anomaly with the time-reversal symmetry, in the $SU(N)$ Yang-Mills theory at $\theta=\pi$. In this appendix, we generalize the discussion to $PSU(N)$ Yang-Mills theory at $\theta= \pi p$ for general $N$ and $p$.\footnote{Similar to the case of the Maxwell theory, there are different versions of $PSU(N)$ Yang-Mills theory. We will focus on the one where every line operator can be either bosonic or fermionic. Such a $PSU(N)$ theory can be defined consistently on spin manifolds with a choice of spin structure.} 
There are two new phenomena: (1) For odd $N$, the non-invertible symmetry of the $PSU(N)$ theory does not arise from a mixed anomaly in the $SU(N)$ theory. (2) For even $N$, the detail of the non-invertible symmetry of the $PSU(N)$ depends on the choice of the counterterm used when gauging the electric $\mathbb{Z}_N^{(1)}$ one-form symmetry.

To start with, we first review the mixed anomaly between the invertible time-reversal symmetry $\mathsf{T}_{\mathsf{SU}}^{\theta=\pi k}$ at $\theta=\pi k$ and the electric $\mathbb{Z}_N^{(1)}$ one-form symmetry in the $SU(N)$ Yang-Mills theory following \cite{Gaiotto:2017yup}. Denote the background gauge field for the electric  $\mathbb{Z}_N^{(1)}$ one-form symmetry by $C\in H^2(X,\mathbb{Z}_N)$, where $X$ is the underlining four-dimensional (spin oriented) manifold. The time-reversal symmetry $\mathsf{T}_{\mathsf{SU}}^{\theta=\pi k}$ can be decomposed into the naive time-reversal transformation $K$,
\ie
&A_0 (t,\vec{x}) \mapsto -A_0 (-t,\vec{x}) \,, \quad A_i (t,\vec{x}) \mapsto A_i (-t,\vec{x}) \,,
\\
&C_{0i} (t,\vec{x}) \mapsto -C_{0i} (-t,\vec{x}) \,, \quad C_{ij} (t,\vec{x}) \mapsto C_{ij} (-t,\vec{x}) \,,
\fe
which maps $\theta=\pi k$ to $\theta=-\pi k$, followed by the  transformation that shifts $\theta\rightarrow\theta+2\pi k$. In the presence of $C$, the time-reversal symmetry $\mathsf{T}_{\mathsf{SU}}^{\theta=\pi k}$ transforms the Lagrangian non-trivially as
\ie\label{eq:SU(N)_anom_TR}
\mathcal{L}^{\mathsf{SU}}_{\pi k}[C]\rightarrow \mathcal{L}^{\mathsf{SU}}_{\pi k}[C]+\frac{2\pi k}{2N}\mathcal{P}(C)~,
\fe
where $\mathcal{P}(C)$ is the Pontryagin square operation that maps $C\in H^2(X,\mathbb{Z}_N)$ to an element in $H^4(X,\mathbb{Z}_{2N})$ when $N$ is even and $\mathcal{P}(C)=C\cup C$ when $N$ is odd. 
This anomalous transformation does not immediately imply an 't Hooft anomaly because we still need to check whether the anomalous transformation can be removed by adding a classical counterterm. Indeed, such counterterm exists when $k$ is even or when $N$ is odd \cite{Gaiotto:2017yup}. Hence, there is a mixed anomaly only when $k$ is odd and $N$ is even.

Now we can gauge the electric $\mathbb{Z}_N^{(1)}$ one-form symmetry in the $SU(N)$ theory to obtain the $PSU(N)$ theory. 
When $N$ is even,  the mixed anomaly between the time-reversal symmetry $\mathsf{T}_{\mathsf{SU}}^{\theta=\pi }$ at $\theta=\pi$ and the electric $\mathbb{Z}_N^{(1)}$ one-form symmetry leads to a non-invertible time-reversal symmetry $\mathsf{T}_{\mathsf{PSU}}^{\theta=\pi }$ in the $PSU(N)$ theory at $\theta=\pi$ after $S$ gauging the electric one-form symmetry \cite{Kaidi:2021xfk}.
How does this generalize to  $\theta=\pi p$? (Recall that $\theta$ is $2\pi N$ periodic in $PSU(N)$ gauge theory.) What happens when $N$ is odd?

Starting from the $SU(N)$ theory, there are multiple ways to construct the $PSU(N)$ theory at $\theta=\pi p$. We can start with the $SU(N)$ theory at $\theta=\pi k$ and perform an $ST^{\frac{k-p}{2}}$ gauging on the $\mathbb{Z}_N^{(1)}$ electric one-form symmetry to obtain the $PSU(N)$ theory at $\theta=\pi p$ if $k-p =0$ mod 2. Here we used the following relation for $PSU(N)$ bundles \cite{Aharony:2013hda, Gaiotto:2017yup,Cordova:2019uob}
\ie
\frac{2\pi}{8\pi^2}\int\text{Tr}(F\wedge F)=-\frac{2\pi}{2N}\int \mathcal{P}(w_2)\text{ mod } 2\pi~,
\fe
where $w_2\in H^2(X,\mathbb{Z}_N)$ is the second Stiefel-Whitney class of the bundle. The $PSU(N)$ theory is then described by the Lagrangian 
\ie
\mathcal{L}_{\pi p}^{\mathsf{PSU}}[B]=\mathcal{L}_{\pi k}^{\mathsf{SU}}[c]+\frac{2\pi}{2N}\frac{k-p}{2}\mathcal{P}(c)+\frac{2\pi}{N}c\cup B~,
\fe
where $c$ is a dynamical $\mathbb{Z}_N$ two-form gauge field and $B\in H^2(X,\mathbb{Z}_N)$ is the background gauge field for the magnetic $\mathbb{Z}_N^{(1)}$ one-form symmetry in the $PSU(N)$ theory. Because of \eqref{eq:SU(N)_anom_TR}, the time-reversal transformation acts on the $PSU(N)$ theory as
\ie\label{eq:trans1}
\mathcal{L}_{\pi p}^{\mathsf{PSU}}[B]\mapsto\mathcal{L}_{\pi k}^{\mathsf{SU}}[c]+\frac{2\pi }{2N}\frac{k+p}{2}\mathcal{P}(c)+\frac{2\pi}{N}c\cup B~.
\fe
Here the time-reversal transformation acts on $B$ as
\ie
B_{0i} (t,\vec{x}) \mapsto B_{0i} (-t,\vec{x}) \,, \quad B_{ij} (t,\vec{x}) \mapsto -B_{ij} (-t,\vec{x})~.
\fe
To undo the time-reversal transformation \eqref{eq:trans1}, we can perform a $(CTST)^p$ transformation on the magnetic $\mathbb{Z}_N^{(1)}$ one-form symmetry. 
Hence, the $PSU(N)$ theory at $\theta=\pi p$ is invariant under the transformation$(CTST)^p K$, which leads to a non-invertible time-reversal symmetry $\mathsf{T}^{\theta=\pi p}_{\mathsf{PSU}}$. The corresponding anti-linear symmetry defects are constructed by performing the above sequence of transformations in half of the spacetime. Following Section \ref{sec:gauging}, the symmetry defects obey a non-invertible fusion rule
\ie
(\mathsf{T}^{\theta=\pi p}_{\mathsf{PSU}})^\dagger\times \mathsf{T}^{\theta=\pi p}_{\mathsf{PSU}}=(\mathcal{C}^{(N)})^p~.
\fe 

Here we note that even though the mixed anomaly is absent in the $SU(N)$ theory when $k$ is even or when $N$ is odd, there can still be a non-invertible time-reversal symmetry in the $PSU(N)$ Yang-Mills theory after gauging. This is because in order to trivialize the anomalous time-reversal transformation \eqref{eq:SU(N)_anom_TR} we need to choose a particular counterterm. 
However, when we gauge the $\mathbb{Z}_N^{(1)}$ electric one-form symmetry to obtain a $PSU(N)$ theory, there are different choices of the counterterm which may not trivialize the anomaly. 
Therefore, the gauging generally leads to a non-invertible time-reversal symmetry in the gauged theory. The details of the non-invertible time-reversal symmetry, such as its fusions, depend on the choice of the counterterm used in the gauging.\footnote{Similar phenomena also appear in \cite{Hsin:2020nts} where the higher-group structure of the symmetry after gauging depends on the choice of the counterterm used in the gauging.} If we choose the same counterterm that trivializes the anomalous time-reversal transformation \eqref{eq:SU(N)_anom_TR}, the theory after gauging is the $PSU(N)$ theory at $\theta\in\pi N\mathbb{Z}$. Indeed, the time-reversal symmetry at these $\theta$-parameters are invertible.

We conclude that a mixed anomaly between the time-reversal symmetry and a $\mathbb{Z}_N^{(1)}$ one-form symmetry leads to a non-invertible time-reversal symmetry after gauging. In some cases, despite the absence of the mixed anomaly, the gauged theory can still have a non-invertible time-reversal symmetry if the counterterm used in the gauging does not trivialize the anomaly and the details of the non-invertible time-reversal symmetry depends on the choice of the counterterm. The same conclusion also holds if we replace the anti-linear time-reversal symmetry by a linear symmetry.

\bibliographystyle{JHEP}
\bibliography{ref}

\end{document}